\documentclass[aps,prd,a4paper,onecolumn,amsmath,showpacs,superscriptaddress,nofootinbib,preprintnumbers,notitlepage]{revtex4-1}
\usepackage{amssymb,amsmath}
\usepackage{graphicx}
\usepackage{color}
\usepackage{pgfplots}
\usepackage{booktabs}
\usepackage{multirow}
\usepackage{dcolumn}
\usepackage{amsmath}
\usepackage{mathtools}
\usepackage{amsfonts}
\usepackage{amssymb}
\usepackage{epstopdf}
\usepackage{bm}
\usepackage{siunitx}
\usepackage{braket}
\usepackage{enumitem}
\usepackage{soul}
\usepackage{ulem}
\usepackage{color}
\usepackage{transparent}
\usepackage{pifont}
\usepackage[font={small}]{caption}
\usepackage{float}
\usepackage{orcidlink}

\newcommand{\be}{\begin{equation}}
\newcommand{\ee}{\end{equation}}
\newcommand{\bea}{\begin{eqnarray}}
\newcommand{\eea}{\end{eqnarray}}

\usepackage{mathrsfs}
\begin{document}
\title{Cosmological dynamics of interacting dark matter-dark energy in generalized Rastall gravity}

\author{Manuel Gonzalez-Espinoza \orcidlink{0000-0003-0961-8029}}
\email{manuel.gonzalez@upla.cl}
\affiliation{Laboratorio de investigaci\'on de C\'omputo de F\'isica, Facultad de Ciencias Naturales y Exactas, Universidad de Playa Ancha, Subida Leopoldo Carvallo 270, Valpara\'iso, Chile}

\author{Ram\'on Herrera \orcidlink{0000-0002-6841-1629}}
\email{ramon.herrera@pucv.cl}
\affiliation{Instituto de F\'{\i}sica, Pontificia Universidad Cat\'{o}lica de Valpara\'{\i}so, Avenida Brasil 2950, Casilla 4059, Valpara\'{\i}so, Chile.
}

\author{Giovanni Otalora \orcidlink{0000-0001-6753-0565}}
\email{giovanni.otalora@academicos.uta.cl}
\affiliation{Departamento de F\'isica, Facultad de Ciencias, Universidad de Tarapac\'a, Casilla 7-D, Arica, Chile}

\author{Carlos  Ríos \orcidlink{0000-0003-3228-0003}}
\email{carlos.rios@ucn.cl}
\affiliation{Departamento de Ense\~nanza de las Ciencias B\'asicas, Universidad Cat\'olica del Norte, Larrondo 1281, Coquimbo, Chile.}

\author{Carlos Rodriguez-Benites \orcidlink{0000-0001-9437-6364}}
\email{cerodriguez@unitru.edu.pe}
\affiliation{Departamento Acad\'emico de F\'{\i}sica, Facultad de Ciencias F\'{\i}sicas y Matem\'aticas, Universidad Nacional de Trujillo, Av. Juan Pablo II s/n, Trujillo, Per\'u}
\affiliation{GRACOCC \& OASIS research groups, Facultad de Ciencias F\'{\i}sicas y Matem\'aticas, Universidad Nacional de Trujillo, Av. Juan Pablo II s/n, Trujillo, Per\'u}

\begin{abstract}
In this work, we investigate late--time interacting cosmologies within the framework of generalized Rastall gravity, where the interaction arises naturally from the non--conservation of the energy--momentum tensor. We formulate the background evolution of the dark sector as an autonomous dynamical system, defining interaction terms $Q_1=\alpha\,\dot{f}$ and $Q_2=-\dot{f}\,(1+\alpha)$, with $\alpha$ a constant  parameter and $f$ a time-dependent function. Three  interaction cases are studied: $f \propto \rho_m$, $f \propto \rho_{de}$, and $f \propto \rho_m + \rho_{de}$, assuming a constant dark-energy equation of state $w_{de}$. For each scenario, we derive the closed dynamical system in terms of the density parameters $(\Omega_{de}, \Omega_m)$, identify its fixed points, and analyze their stability across the parameter space. In this context, the phase-space exhibits  a standard cosmological   dynamics: an unstable radiation point, a transient matter saddle, and a stable late--time attractor with accelerated expansion. 
 In addition, we utilize a joint likelihood analysis with Cosmic Chronometers, PantheonPlus, and DESI data to obtain marginalized parameter estimates at the 68$\%$ and 95$\%$ confidence levels, constraining  the  parameter space in each interaction model.

\end{abstract}

\maketitle
\section{Introduction}
The late-time acceleration of the Universe, robustly supported by a variety of observational evidence, has motivated a broad exploration of scenarios in which  dark energy is not merely a passive cosmological constant but rather a dynamical component possibly coupled to dark matter. Within the wide range of models proposed to describe the current universe, special attention has been given to those that incorporate interactions between their components \cite{Wang:2016lxa,Zimdahl:2001ar,Chimento:2003iea}. In this framework,
 models of interacting dark energy and dark matter (DE-DM) present a systematic approach to modifying both the background expansion and the growth of structures, while simultaneously addressing questions of coincidence without invoking exotic instabilities.  With respect  to the analysis of  the dynamics evolution of these interacting models, the background  can be studied into autonomous dynamical systems and analyzed in relation to model parameters. In this respect, the phase-space studies in different cosmological models have clarified how compactification techniques, critical-point classification and stability criteria provide a robust qualitative control over entire families of scenarios \cite{PhaseSpaceMethods2025,Gonzalez-Espinoza:2025bzd}. 
Interacting dark sector model has been extensively studied over the last decade, showing that even  simplest interaction can qualitatively modify the different-time trajectories, trigger scaling regimes, during different epochs in the evolution of the Universe \cite{Copeland:2006wr,Copeland:1997et,delCampo:2005tr,Amendola:1999qq,delCampo:2013hka}. For instance, in interacting vector-like dark energy (IVDE), linear and non-linear couplings lead to distinct fixed points and to sign-changeable energy transfer along the history, with concrete predictions for $w_{de}(z)$ and $H(z)$. These results illustrate how interaction affects both the background and thermodynamic consistency of the accelerating universe and motivate a careful mapping of critical points in any new interacting setup \cite{cid2021bayesian, rodriguez2020universe, rodriguez2024revisiting, VectorLikeDE2025}. Relatedly, in torsion-based modified gravity with energy and momentum exchange (e.g., non-minimal torsion--matter couplings modeled via Sorkin--Schutz fluids), rigorous dynamical analyses uncover scaling regimes and accelerated attractors, underscoring the value of placing interactions within a fully specified gravitational theory before drawing phenomenological conclusions \cite{ gonzalez2018growth,ShabaniMoradpourZiaie2022_GRG_DS}. For a review  of different interacting models during at the  present epoch, see Refs.\cite{delCampo:2015vha,Farrar:2003uw,Valiviita:2008iv,delCampo:2008jx,delCampo:2006vv}.

In relation to the cosmological  models  based on modified gravity, we should mention those that utilize the Rastall gravity \cite{Rastall:1972swe}. 
Rastall’s model of gravity  provides an alternative theoretical framework for exploring the dynamics of the early and present universe \cite{Rastall:1972swe}. This theory adopts a phenomenological approach to encapsulate the covariant effects of quantum fields in curved spacetime \cite{Bertlmann:1996xk}. Although it does not possess a fundamental action, effective formulations have been suggested by modifying the Einstein–Hilbert action, frequently within the context of $f(R,T)$ gravity~\cite{DeMoraes:2019mef,Shabani:2020wja}. However, the original model faces considerable challenges in being integrated into a conventional Lagrangian framework \cite{Fabris:2020uey}. In cosmology, Rastall gravity has been investigated both in the early universe, where scalar-field dynamics modified by energy–momentum non-conservation can align with observational constraints on inflation~\cite{Batista:2011nu,Afshar:2023uyw,Mandal:2023ink,Herrera:2025duu}, and in later epochs, where it can emulate dark energy behavior, resembling quintessence or phantom regimes contingent on the value of the Rastall parameter \cite{Fabris:2017msx,Singh:2024urv,Singh:2024urv}. A generalized extension, wherein the Rastall parameter is dynamic, has also been proposed; this results in non-singular emergent scenarios and establishes connections with particle creation in Einstein gravity \cite{Moradpour:2017shy}. Consequently, a broad spectrum of cosmological models has been explored in the literature within this generalized framework (see, e.g., Refs.\cite{Lin:2018dgx,Lin:2020fue,Ziaie:2020ord,Shabani:2022zlx,Das:2018dzp}). In Rastall’s formulation, the standard divergence-free condition of the energy–momentum tensor is relaxed, permitting a geometry-mediated exchange between matter and the gravitational sector while maintaining overall covariance. This modification alters the Friedmann equations and provides a platform for examining background interaction terms and their dynamic role in phase space. Within this broader perspective, generalized Rastall-type models have been utilized to capture diverse cosmological behaviors (matter- and radiation- dominated eras as well as accelerated expansion) and to naturally connect with autonomous dynamical formulations suitable for fixed-point and bifurcation analyses, which we adopt as the methodological foundation of this work \cite{Moradpour2017_EPJC_RastallGeneralization,ShabaniMoradpourZiaie2022_GRG_DS}. Beyond strictly phenomenological couplings, non-minimal matter--geometry interactions have been explored as a conceptual underpinning for effective $Q$-terms. Curvature-matter couplings generate modified conservation laws and extra force terms in the motion of matter fields, offering a Lagrangian route to interaction kernels and clarifying consistency requirements. This broader context helps situate Rastall gravity among theories where exchange between geometry and matter is not ad hoc but structurally permitted \cite{BertolamiParamos2008_NMCmatter, BertolamiEtAl2007_PRD_ExtraForce}. In parallel, creation of particles by gravity provides an effective fluid description in which a negative creation pressure can mimic $w_{\rm eff}<-1$, another conceptual bridge to interaction terms that we keep in view when interpreting effective equations of state \cite{NunesPavon2015_PRD_ParticleCreation}. We will also note analogies with other modified frameworks (e.g., Ref.\cite{KiritsisKofinas2009_HLCosmology}) where departures from the General Relativity (GR) reorganize background dynamics and can be cast in autonomous form, reinforcing the value of the dynamical-systems diagnostics we deploy here \cite{Mukohyama2010_HLCosmologyReview}.

In this article, we explore the cosmological dynamics of the interaction between dark energy and dark matter within the framework of generalized Rastall gravity. Within this framework, we analyze  how the background dynamics affect the different epochs of the Universe for various interaction terms. Thus,  we
 examine three scenarios that emerge either from an interaction function $f(t)$, or equivalently, through the source terms defined as  $Q_1=\alpha\dot{f}$ and  $Q_2=-(Q_1+\dot{f})$, where $\alpha$ represents a coupling parameter. For the analysis of the DE--DM interaction dynamics, we consider three functional forms of the coupling function $f=f(t)$ associated with the different energy densities of the dark sector: $f=\beta \rho_m$, $f=\beta\rho_{de}$, and $f=\beta\left(\rho_m+\rho_{de}\right)$, with $\beta$ being a proportionality constant. Working at the background level, we recast the modified Friedmann and continuity equations into a closed autonomous system, construct the critical sets (including matter, and DE, dominated candidates and possible scaling solutions), and perform linear stability analysis to chart the late-time fates (accelerated attractors, saddles, and non-accelerating points) across parameter space. Our strategy mirrors established dynamical scenarios (compactification of non-compact variables, careful treatment of singular submanifolds, and, when needed, center-manifold or numerical checks), ensuring comparability with prior interacting models and with generalized Rastall formulations \cite{PhaseSpaceMethods2025,ShabaniMoradpourZiaie2022_GRG_DS}.

The Rastall framework inherently permits deviations from the conventional conservation law, allowing the interaction terms to be interpreted as geometry-induced exchanges rather than insertions from external sources. This perspective facilitates the differentiation between genuine physical couplings and mere reparametrizations, while elucidating the specific conditions under which General Relativity is restored, as explicitly reflected in the choice of dynamical variables. Furthermore, a comparative analysis of the interaction forms $Q_1$ and $Q_2$ reveals that the location and stability of the accelerating fixed points, as well as the emergence of scaling accelerator solutions, depend not only on the parameters $\alpha$ and $\beta$, but also on the Rastall parameter governing non-conservation. This interaction, absent in minimally coupled RG baselines, underlines the distinctive role of the Rastall modification~\cite{Rastall:1972swe,Moradpour2017_EPJC_RastallGeneralization}.

The main contributions of this study are as follows. We develop a comprehensive autonomous framework for the interaction between dark energy and dark matter within the context of generalized Rastall gravity, and we delineate the global phase-space structure for various interaction scenarios characterized by distinct $f$ functions. In addition, we establish conditions for the existence and stability of accelerating critical points, offering insights into the scaling possibilities and their sensitivity to model parameters. Furthermore, we examine effective-fluid interpretations, including analogies with creation pressure, and investigate the correspondence with other interaction frameworks, such as vector-like DE and scalar–torsion couplings. This analysis highlights both the shared dynamical features and the significant differences arising from Rastall's non-conservation law~\cite{VectorLikeDE2025,gonzalez2018growth, NunesPavon2015_PRD_ParticleCreation}.

In addition,  to constrain the different free parameters of the interacting   model, we implement a joint likelihood analysis based on three independent cosmological probes: Cosmic Chronometers, Type Ia Supernovae , and Baryon Acoustic Oscillations  from the DESI survey. In this form, we determine that  
each dataset provides  supplementary  information on the cosmic expansion history and allows us to break parameter degeneracies.

The structure of this work is organized as follows: in the subsequent section, we present a brief overview of the fundamental equations within the Rastall framework, along with the background equations that include interaction terms. In Section \ref{DS} we study the dynamical system for our model, in the framework of  Rastall gravity. Here, we introduce dimensionless variables and formulate the corresponding autonomous system. Also,  we analyze the dynamics of the several parameters associated with different interacting $f(t)$-model together with the critical points and their stability for the interaction forms $Q_1$ and $Q_2$.  In Section \ref{ST} we present the statistical analysis on the basis of  $\chi^2$ estimators, to constrain the different parameters of each interacting model. Finally, our conclusions are presented \ref{conclusion}. In this work we chose units such that
$c=\hbar=1$.

\section{Generalized Rastall Theory}

The generalized Rastall theory is characterized by a modification of Einstein’s field equations, where in the covariant conservation of the energy-momentum tensor $T_{\mu\nu}$ is relaxed. In this context, following Refs.\cite{Moradpour:2017shy,Das:2018dzp}, a deviation from the standard conservation law can be introduced, which takes the form of the following non-conservation equation for the energy–momentum tensor.
\begin{equation}
\label{noncT}
\nabla_\mu T^{\mu\nu}= \nabla^\nu \left[\frac{\left(1-\lambda_\text{Ras}\right)}{2\kappa}R\right],
\end{equation}
where $R$ corresponds to the Ricci scalar and the function $\kappa$ must be determined in the weak field limit, resulting in \cite{Rastall:1972swe}
\begin{equation}
 \kappa= \kappa^{\text{(GR)}}\left(\frac{2\lambda_\text{Ras}-1}{3\lambda_\text{Ras}-2}\right). 
\end{equation}
Here, the quantity $\kappa^{\text{(GR)}}=8\pi G=8\pi/M_p^2$ corresponds to the Einstein coupling constant in general relativity (GR), and here $M_p$ is the Planck mass. It can be seen that when $\lambda_\text{Ras}$ is a constant equal to one, the GR is restored. As a result, the traditional conservation law $\nabla_\mu T^{\mu\nu}= 0$ is also reestablished. 

Within this framework, the constant parameter of the original Rastall theory \cite{Rastall:1972swe} is replaced by a space-time-dependent function $\lambda_\text{Ras}$, thereby introducing a dynamical coupling between geometry and matter. In the generalized Rastall gravity framework, the Rastall parameter $\lambda_\text{Ras}$, which in the standard formulation is a constant, is promoted to a dynamic quantity depending on the space-time coordinates, i.e. $\lambda_\text{Ras}=\lambda_\text{Ras}(x^\mu)$. In particular, we assume in the following that the variation of $\lambda_{\text{Ras}}$ is homogeneous on cosmological scales, allowing it to be treated as a purely time-dependent function.

The modified field equations that characterize the generalized Rastall framework are expressed as follows \cite{Moradpour:2017shy}
\begin{equation}
\label{campoRas}
    R_{\mu\nu} - \frac{1}{2}g_{\mu\nu} \left(R\lambda_{\text{Ras}}\right) = \kappa  T_{\mu\nu},
\end{equation}
where $R_{\mu\nu}$ corresponds to the Ricci tensor and $g_{\mu\nu}$ is the metric tensor.

In relation to matter, the effective energy–momentum tensor can be reformulated in the form of a perfect fluid energy–momentum tensor, characterized by a total energy density $\rho_{\text{tot}}$ and a total pressure $p_{\text{tot}}$ associated with the matter content of the Universe. In this way, setting $\nu=0$ in Eq.(\ref{noncT}), we explicitly derive the non-conservation equation within the framework of the Rastall theory, obtaining:
\begin{equation}
\dot{\rho}_\text{tot}+3H(\rho_\text{tot}+p_\text{tot})=-\frac{d}{dt}\left[\frac{\left(1-\lambda_\text{Ras}\right)}{2\kappa}R\right]\label{rhotot1},
\end{equation}
where $H=\dot{a}/a$ is the so-called Hubble parameter, $a(t)$ is the scale factor, and $t$ is the cosmic time. In addition, in the following we denote derivatives with respect to cosmic time $t$ by dots.

By introducing a time-dependent function $f(t)$ defined by
\begin{equation}
\frac{1-\lambda_\text{Ras}}{2\kappa} = \frac{f(t)}{R}, \label{ft}
\end{equation}
we observe that Eq.(\ref{rhotot1}) can be rewritten in a reformulated form.
 Furthermore, in the special case where $f(t) = 0$, the parameter $\lambda_\text{Ras}$ takes the value 1, thus reducing the theory to standard GR.
 
 Thus, by considering Eq.(\ref{ft}), we can rewrite Eq.(\ref{rhotot1}) as
\begin{equation}
\dot{\rho}_\text{tot}+3H(\rho_\text{tot}+p_\text{tot})=-\dot{f}(t),\label{rhotot2}
\end{equation}
where the function $\dot{f}(t)$ accounts for the non conservation in Rastall gravity.

With regard to the Friedmann equation, we consider a spatially flat Friedmann–Robertson–Walker (FRW) metric in Eq.(\ref{campoRas}), which then yields:
\begin{equation}
(2\lambda_\text{Ras}-1)H^2+(\lambda_\text{Ras}-1)\dot{H}=\frac{\kappa}{3}\rho_\text{tot}=\frac{\kappa}{3}\left(\rho_{de}+\rho_m+\rho_r\right),  \label{Feq1}
\end{equation}
where we have considered the total energy density $\rho_{tot}=\rho_{de}+\rho_{m}+\rho_r$. Here, $\rho_{de}$ corresponds to the energy density associated with dark energy, and $\rho_m$ and $\rho_r$ denote the energy densities of matter and radiation, respectively. 

In addition, the second Friedmann equation, also known as the Raychaudhuri equation, is represented by 
\begin{equation}
 \frac{\ddot{a}}{a}=\frac{1}{(3\lambda_\text{Ras}-1)}\left[(2-3\lambda_\text{Ras})H^2-\kappa(\lambda_\text{Ras}) p_\text{tot}\right].\label{Feq2} \end{equation}
It can be observed that by setting $\lambda_\text{Ras}=1$ in Eqs.(\ref{Feq1}) and (\ref{Feq2}), these equations reduce to the standard Friedmann equations. 

To investigate the interaction between dark matter and dark energy, we assume that these components are coupled through two source terms, denoted by 
$Q_1$ and 
$Q_2$, respectively. These terms enter the corresponding energy balance equations of dark matter and dark energy, respectively. Accordingly, the equations of motion for the interacting dark sector from Eq.(\ref{rhotot2}) can be written as
\begin{equation}
\dot{\rho}_m+3H\rho_m=\alpha\dot{f}=Q_1,\label{Q1}
\end{equation}
and
\begin{equation}
\dot{\rho}_{de}+3H\rho_{de}(1+w_{de})=-\dot{f}(1+\alpha)=Q_2=-\,\left(Q_1+\dot{f}\,\right),\label{Q2}
\end{equation}
where $\alpha$ is a dimensionless parameter and $w_{de}$ corresponds to the equation of state associated with the dark energy. In addition, we note that the interaction terms are constrained by $Q_1 + Q_2 = -\dot{f}$, which encodes the departure from energy–momentum conservation in this framework.

In addition, for radiation, the equation of motion becomes
\begin{equation}
\dot{\rho}_r+4H\rho_r=0.
\end{equation}

As in general relativity, the dimensionless density parameters $\Omega_i(t)$ can be defined to represent the fractional contribution of each component of the Universe $i$ (radiation, dark matter, or dark energy) to the critical energy density $\rho_c$. Within this generalized Rastall framework, the introduction of a time-dependent coupling and a modified gravitational sector requires the incorporation of a critical energy density denoted $\rho_c$. In this context, utilizing Eqs.(\ref{Feq1}) and (\ref{Feq2}), together with the relation $\ddot{a}a^{-1}=\dot{H}+H^2$, we derive the critical energy density within the Einstein frame as follows:
\begin{equation}
 \rho_c=\frac{3H^2}{\kappa^\text{(GR)}}=\frac{1}{2(3\lambda_\text{Ras}-2)}\left[(3\lambda_\text{Ras}-1)\rho_\text{tot}+3(\lambda_\text{Ras}-1)p_\text{tot}\right]\label{rhoceff}.
\end{equation}

On the other hand, introducing the relation between total pressure and total energy density, given by  $p_\text{tot}=\omega_\text{tot}\,\rho_\text{tot}$, where $\omega_\text{tot}$ denotes the  total equation of state (EoS) parameter defined as 
\begin{equation}
    \omega_\text{tot}=\frac{\sum(\omega_i\rho_i)}{\sum\rho_i}=\frac{p_{de}+p_r}{\rho_m+\rho_{de}+\rho_r}=\frac{w_{de}\rho_{de}+w_r\rho_r}{\rho_m+\rho_{de}+\rho_r}=\frac{w_{de}\rho_{de}+\rho_r/3}{\rho_m+\rho_{de}+\rho_r}\label{wt}
\,\,,\,
\end{equation}
where we have used the relations $p_{de} = w_{de}\rho_{de}$ and $p_r = w_r\rho_r$, with $w_r = 1/3$ denoting the EoS parameter for radiation. In relation to the EoS parameter for dark energy $w_{de}$, in the following we will assume that $w_{de}$ is constant.

Under these considerations Eq.(\ref{rhoceff}) can be rewritten as 
\begin{equation}
 \rho_c=\frac{3H^2}{\kappa^\text{(GR)}}=g(\lambda_\text{Ras},w_{de},\Omega_i)\rho_\text{tot}=g(\lambda_\text{Ras},w_{de},\Omega_{i})[\rho_r+\rho_{de}+\rho_m],\label{H1}
\end{equation}
where the function $g(\lambda_\text{Ras},w_{de}, \Omega_i)$ depends on the EoS parameter associated with the dark energy, the Rastall parameter, and on the density parameter of each fluid component, $\Omega_i$, defined as $\Omega_i=\rho_i\,\kappa^\text{(GR)}/3H^2$, such that

\begin{equation}
g(\lambda_\text{Ras},w_{de}, \Omega_i)=g=\,\frac{1}{2(3\lambda_\text{Ras}-2)}\left[(3\lambda_\text{Ras}-1)+3(\lambda_\text{Ras}-1)\,\omega_\text{tot}\right].\label{ga}
\end{equation}
Introducing the previously mentioned dimensionless density parameters associated with each fluid, $\Omega_i$, we have
\begin{equation}
\Omega_{de}=\frac{\rho_{de}\kappa^\text{(GR)}}{3H^2}, \,\,\,\,\Omega_r=\frac{\rho_{r}\kappa^\text{(GR)}}{3H^2}\,\,\,\,\,\mbox{and}\,\,\,\,\,\Omega_m=\frac{\rho_{m}\kappa^\text{(GR)}}{3H^2}.
\end{equation}
Thus, we determine that the total effective equation of state parameter, $\omega_\text{tot}$, as expressed in Eq.(\ref{wt}), can be rewritten as follows
\begin{equation}
\omega_\text{tot}=\frac{w_{de}\Omega_{de}+w_r\Omega_{r}}{\Omega_{de}+\Omega_r+\Omega_{m}}=\frac{w_{de}\Omega_{de}+\Omega_{r}/3}{\Omega_{de}+\Omega_r+\Omega_{m}}=\left(w_{de}\Omega_{de}+\Omega_{r}/3\right)\,g,
\end{equation}
where we have used from Eq.(\ref{H1}) that $\Omega_{de}+\Omega_m+\Omega_r =1/ g$. 

In relation to the Rastall parameter and its variation with respect to time, by combining Eqs.(\ref{ft}) and (\ref{Feq1}), and considering that the Ricci scalar for a flat FRW metric is given by  $R=6[\ddot{a}/a+H^2]=6[\dot{H}+2H^2]$, we find that the Rastall parameter can be expressed as
\begin{equation}
\lambda_\text{Ras}=\left[\frac{2f+(1-3w_{de})\rho_{de}+\rho_m}{4f+(1-3w_{de})\rho_{de}+\rho_m}\right]=\left[\frac{2f\kappa^{(GR)}/3H^2+(1-3w_{de})\Omega_{de}+\Omega_m}{4f\kappa^{(GR)}/3H^2+(1-3w_{de})\Omega_{de}+\Omega_m}\right].\label{Ras}
\end{equation}
Here, we note that in the particular case where there is no interaction, the function $f=0$, and Eq.(\ref{Ras}) reduces to $\lambda_\text{Ras}\rightarrow$1.

In addition, using the aforementioned relationships, we find that the function
$g(\lambda_\text{Ras},w_{de},\Omega_i)$ given by Eq.(\ref{ga}) depends only on the interaction term $f$, $w_{de}$ and $\Omega_i$, and yields
\begin{equation}
g(f,w_{de}, \Omega_i)=g=\,\left[\frac{(3\lambda_\text{Ras}-1)}{2(3\lambda_\text{Ras}-2)-(\lambda_\text{Ras}-1)[\Omega_r+3w_{de}\Omega_{de}]}\right],
\label{GA}
\end{equation}
with $\lambda_\text{Ras}$ given by Eq.(\ref{Ras}).
As before, in particular, we note that in the GR limit, where the Rastall parameter $\lambda_{\text{Ras}} = 1$, Eq.(\ref{GA}), which involves the function $g$, reduces to $g \rightarrow 1$.

It is convenient to introduce dimensionless density parameters, defined as
\begin{equation}
\bar{\Omega}_{de}=g\,\Omega_{de},\,\,\,\,\,\,\,\bar{\Omega}_{m}=g\,\Omega_{m},\,\,\,\,\,\,\,\mbox{and}\,\,\,\,\,\,\,\,\,\bar{\Omega}_{r}=g\,\Omega_{r},\label{bO}
\end{equation}
respectively. Thus, from these definitions and using Eq.(\ref{H1}), we obtain the following constraint equation:
\begin{equation}
\bar{\Omega}_{\text{tot}}(t)=g\,\sum_i\Omega_i(t)=g\,\left(\Omega_{de}+\Omega_r+\Omega_{m}\right)=\bar{\Omega}_{de}+\bar{\Omega}_r+\bar{\Omega}_{m} =\,1,
\end{equation}
where the function $g$ is defined by the Eq.(\ref{GA}). 

In the following, we describe the dynamical system for different interaction models within the framework of generalized Rastall gravity.

\section{Dynamical system}\label{DS}

In this section, we analyze the dynamical system of our different interaction models in the generalized Rastall gravity, to determine its critical points, together with the corresponding cosmological parameters and the stability of the autonomous system. 

In the following, we examine three types of interactions which emerge either from the interaction function $\dot{f}$ or from the source terms $Q_1=\alpha\,\dot{f}$ and $Q_2=-\dot{f}(1+\alpha)$. We then provide a detailed dynamical analysis of these interaction models.

Examining the left-hand sides of Eqs.(\ref{Q1}) and (\ref{Q2}), it is evident that the interaction function $f$ must depend on the energy densities $\rho_{de}$ or $\rho_{m}$, see Refs.\cite{Perez:2021cvg,Shahalam:2015sja,Choudhury:2018vat}. In the literature, various combinations of these quantities have been considered, so that the interaction function $f$ generally takes the form $f = f( \rho_m ,  \rho_{de},  \rho_m+\rho_{de}, \rho_{m}-\rho_{de})$, such that the interaction term $\dot{f} = \dot{f}( \dot{\rho}_m ,  \dot{\rho}_{de},  \dot{\rho}_m+\dot{\rho}_{de},\dot{ \rho}_{m}-\dot{\rho}_{de})$. In this work, we focus on the three simplest functions $f(t)$ that account for the interaction models, which are
\begin{equation}
f(t)=\beta\,\rho_{m},\,\,\,\,\,\,\,\,f(t)=\beta\,\rho_{de},\,\,\,\,\,\,\,\,\,\,\mbox{and}\,\,\,\,\,\,\,\,\,f(t)=\beta\,(\rho_{m}+\rho_{de}),
\end{equation}
where $\beta$ is a dimensionless constant parameter. Thus, the source terms $Q_1$ and $Q_2$, associated with the three simplest interaction models within the framework of generalized Rastall gravity, are given by the following expressions:
\begin{equation}
Q_1=\alpha\beta\dot{\rho}_m\,\,\,\,\,\,\,\,\mbox{and}\,\,\,\,\,\,\,Q_2=- \beta\dot{\rho}_m(1+\alpha),\label{q1}
\end{equation}
\begin{equation}
Q_1=\alpha\beta\dot{\rho}_{de}\,\,\,\,\,\,\,\,\mbox{and}\,\,\,\,\,\,\,\,Q_2=- \beta\dot{\rho}_{de}(1+\alpha),\label{q2}
\end{equation}
and
\begin{equation}
Q_1=\alpha\beta(\dot{\rho}_{m}+\dot{\rho}_{de})\,\,\,\,\,\,\,\,\,\,\,\mbox{and}\,\,\,\,\,\,\,\,\,Q_2=- \beta(\dot{\rho}_{m}+\dot{\rho}_{de})\,(1+\alpha).\label{q3}
\end{equation}



In the following, we analyze the critical points and their stability together with the  $\chi^2$ test, for the different interaction models given by Eqs.(\ref{q1}), (\ref{q2}) and (\ref{q3}), within the framework of generalized Rastall gravity.

\subsection{Interaction function 
$f(t) = \beta \rho_m(t)$. Critical points and Stability of critical points.}

In this section, we present the dynamical system corresponding to the first interaction function, defined as $f=\beta\,\rho_m$, together with the coupling function expressed in terms of the scale factor. Within this framework, the dynamical system can be written in a general form, and its evolution is governed by the following equations:
\begin{eqnarray}
\dfrac{d \Omega _{de}}{d N} &=& \dfrac{1}{(\alpha  \beta -1) \left[\left(3 w_{de}-1\right) \Omega _{de}+(2 \beta -1) \Omega _m\right]} f_1 (\Omega _{de}, \Omega _{m}), 
\label{dinsyseq1}\\
\dfrac{d \Omega _{m}}{d N} &=& \dfrac{1}{(\alpha  \beta -1) \left[\left(3 w_{de}-1\right) \Omega _{de}+(2 \beta -1) \Omega _m\right]} f_2 (\Omega _{de}, \Omega _{m}), 
\label{dinsyseq2}
\end{eqnarray}
where $N$ denotes the number of $e-$folds and  is defined in terms of the scalar factor ‘‘$a$’’ as $N=\ln\,a$. In addition,  the functions $f_i(\Omega_{de},\Omega_m)$, with $i=1,2$ are explicitly given by
\begin{eqnarray}
f_1 (\Omega _{de}, \Omega _{m}) &=& \Omega _{de} \Omega _m \left(\beta -3 (\beta +1) (2 \alpha  \beta +1) w_{de}+2 \beta ^2 \left(\alpha +2 \alpha  \Omega _m\right)+\alpha  \beta  \left(\Omega _m+2\right)-4 \beta  \Omega _m-\Omega _m+1\right) \nonumber\\
&&-(\alpha  \beta -1) \left(3 w_{de}-1\right) \Omega _{de}^2 \left(3 w_{de}+(4 \beta +2) \Omega _m-1\right)+(\alpha  \beta -1) \left(1-3 w_{de}\right){}^2 \Omega _{de}^3\nonumber\\
&&-3 (\alpha +1) \beta  (2 \beta -1) \Omega _m^2,\label{ff1} \nonumber\\ 
&&\\
f_2 (\Omega _{de}, \Omega _{m}) &=& \Omega _m (-\left(3 w_{de}-1\right) \Omega _{de} \left(2 \alpha  \beta  \left(2 \beta  \Omega _m+\Omega _m-2\right)-2 (2 \beta +1) \Omega _m+1\right)+(\alpha  \beta -1) \left(1-3 w_{de}\right){}^2 \Omega _{de}^2 \nonumber\\
&&+\Omega _m \left(4 \alpha  \beta ^2 \left(\Omega _m+2\right)+\beta  \left(\alpha  \left(\Omega _m-4\right)-4 \Omega _m-2\right)-\Omega _m+1\right)\big). 
\label{ff2}
\end{eqnarray}

Additionally, from Eq.(\ref{GA}) we find that the function $g$ in terms of the dynamical variables for this interaction model results in the following: 
\begin{eqnarray}
    g(\Omega _{de}, \Omega _{m}) = g=\frac{-3 w_{de} \Omega _{de}+\Omega _{de}+\Omega _m}{\left(3 w_{de}-1\right) \Omega _{de} \left(\beta  \Omega _m-1\right)-\beta  \left(\Omega _m+2\right) \Omega _m+\Omega _m},\label{GG1}
\end{eqnarray}
and the Rastall parameter for our interaction model from Eq.(\ref{Ras}) becomes
\begin{equation}
\lambda_\text{Ras}(\Omega_{de},\Omega_m)=\lambda_\text{Ras}=\left[\frac{(2\beta+1)\Omega_m+(1-3w_{de})\Omega_{de}}{(4\beta+1)\Omega_m+(1-3w_{de})\Omega_{de}}\right].\label{Ras1}
\end{equation}

Using these functions $f_i(\Omega_{de},\Omega_m)$, we will determine the critical points corresponding to our first interaction term. These points are obtained by imposing the conditions;
$d\Omega_{de}/dN=d\Omega_{m}/dN=0$ in Eqs.(\ref{dinsyseq1}) and (\ref{dinsyseq2}), respectively. Moreover, taking into account the definitions of the dynamical variables $\Omega_{de}$ and $\Omega_m$, the physically meaningful values at the critical points must satisfy conditions
$\Omega_{{de}_c}\ge 0$ and $\Omega_{{m}_c}\ge 0$, respectively. In the following, the subscript ``$c$'' indicates a critical point. Moreover, Table \ref{table1} presents the critical points of the system associated with the first interaction function $f=\beta\rho_m$, in the framework of generalized Rastall gravity. Furthermore, Table \ref{table2} shows the corresponding values of their cosmological parameters. Here, the quantities $\bar{\Omega}_{de}$ and $\bar{\Omega}_{m}$ are defined by Eq.(\ref{bO}).

In these tables, $a_R$
refers to the critical point corresponding to a radiation epoch. At this critical point, the radiation parameter is $\bar{\Omega}_r=1$
and a total EoS parameter
$\omega_{tot}=1/3$ . This point is independent of the value  $\alpha$. Moreover, we also note that it does not depend on the interaction parameter 
$\beta$.

The critical point $b_M$ is obtained only when the coupling takes the special value $\alpha=-1$. In that case, matter fully dominates the dynamics, with $\bar{\Omega}_m=1$ 
and a total EoS $\omega_{tot}=0$, as expected for a pressureless fluid. Thus,  in order to obtain a matter-dominated epoch, our model requires setting the coupling parameter to $\alpha = -1$. This condition implies that the model features an interaction only with dark matter, since the interaction source associated with dark energy vanishes, i.e., $Q_2 = 0$, (see Eqs.(\ref{Q2})). Here, the interaction present only in the matter equation of motion, and not in the dark energy component, can be associated with particle production. This class of scenarios was investigated in Refs.\cite{Prigogine:1989zz,NunesPavon2015_PRD_ParticleCreation,Parker:1968mv,Parker:1971pt,Calvao:1991wg,Pramanik:2025jwe}, where a consistent framework was developed to incorporate matter creation within Einstein’s field equations. The key idea was to modify the standard particle number balance equation by adding a source term on the right-hand side, thereby accounting for the continuous production of particles. In addition, the same state in which $\bar{\Omega}_m=1$ also appears when $\beta=0$, but it simply corresponds to the standard matter epoch of the background model. As the interaction vanishes in the limit $\beta = 0$, we exclude this case from the analysis of interacting scenarios.

The critical point $c$ represents a dark energy–dominated epoch, where the cosmological parameter associated with dark energy is $\bar{\Omega}_{de}=1$. At this point, the total EoS parameter coincides with the  EoS parameter related to the dark energy $w_{de}$.

On the other hand, to investigate the stability of the critical points, we introduce small time-dependent linear perturbations in the dimensionless variables of the dynamical system around each critical point. Specifically, we consider $\Omega_{de}=\Omega_{{de}_c}+\delta\Omega_{de}$ and $\Omega_{m}=\Omega_{{m}_c}+\delta\Omega_{m}$, where $\delta\Omega_{de}$ and $\delta\Omega_{m}$
denote small perturbations with respect to the background values, such that these perturbations satisfy; $|\delta\Omega_{de}|\ll 1$ and $|\delta\Omega_{m}|\ll 1$, respectively.
Substituting these perturbations into Eqs.(\ref{dinsyseq1}) and (\ref{dinsyseq2}) yields the linear perturbation matrix $\mathcal{M}$ according to Ref.\cite{Copeland:2006wr}. The stability of each critical point is then analyzed through the eigenvalues of $\mathcal{M}$, denoted by $\mu_i$ ($i=1,2$), which are evaluated at the corresponding fixed point.

Regarding the classification of the stability of critical points, a point is identified as a stable node if all eigenvalues are negative, while it is an unstable node if all eigenvalues are positive. A saddle point arises when the eigenvalues exhibit different signs. Moreover, the system describes a stable spiral when the determinant of the matrix $\mathcal{M}$ is negative. It is worth noting that critical points classified as stable nodes or stable spirals act as attractors in the phase space.

In what follows, we provide a detailed examination of the eigenvalues corresponding to the critical points of the proposed interaction model within the framework of generalized Rastall gravity, along with the conditions that determine their stability.

\begin{figure}[!tbp]
  \centering
    \includegraphics[width=0.5\linewidth]{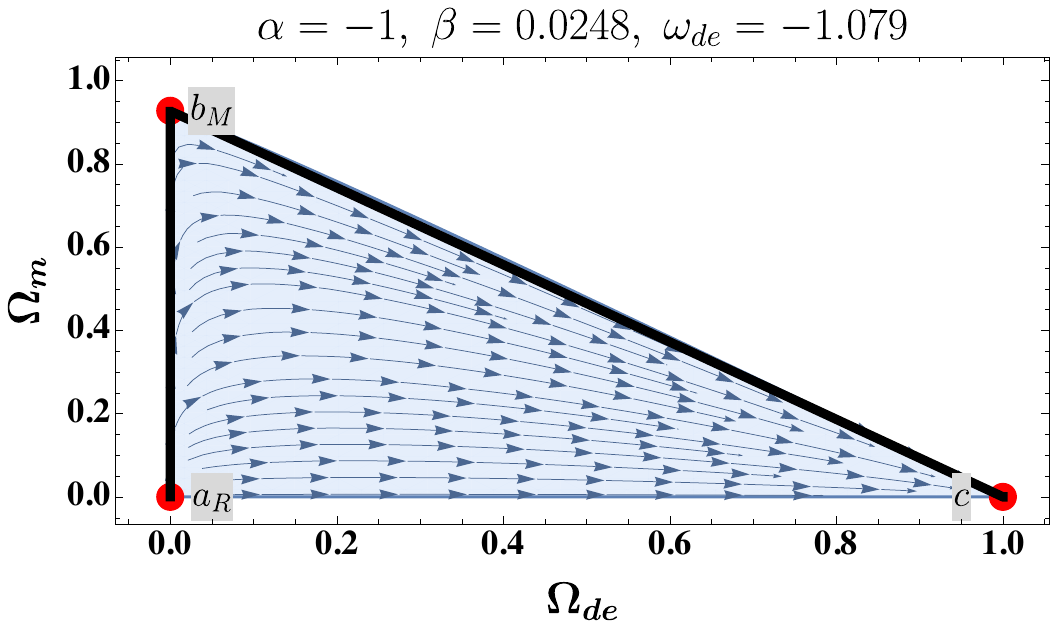} 
    \caption{\scriptsize{This figure depicts the phase-space evolution $(\Omega_m,\Omega_{de})$ for the first interaction function within generalized Rastall gravity, with parameters given by the mean values of Table \ref{mean_inter1} for \textbf{CC + PantheonPlus  + DESI}.
    }
    } 
    \label{Fig1}
\end{figure}

\begin{figure}[!tbp]
  \centering
    \includegraphics[width=0.5\linewidth]{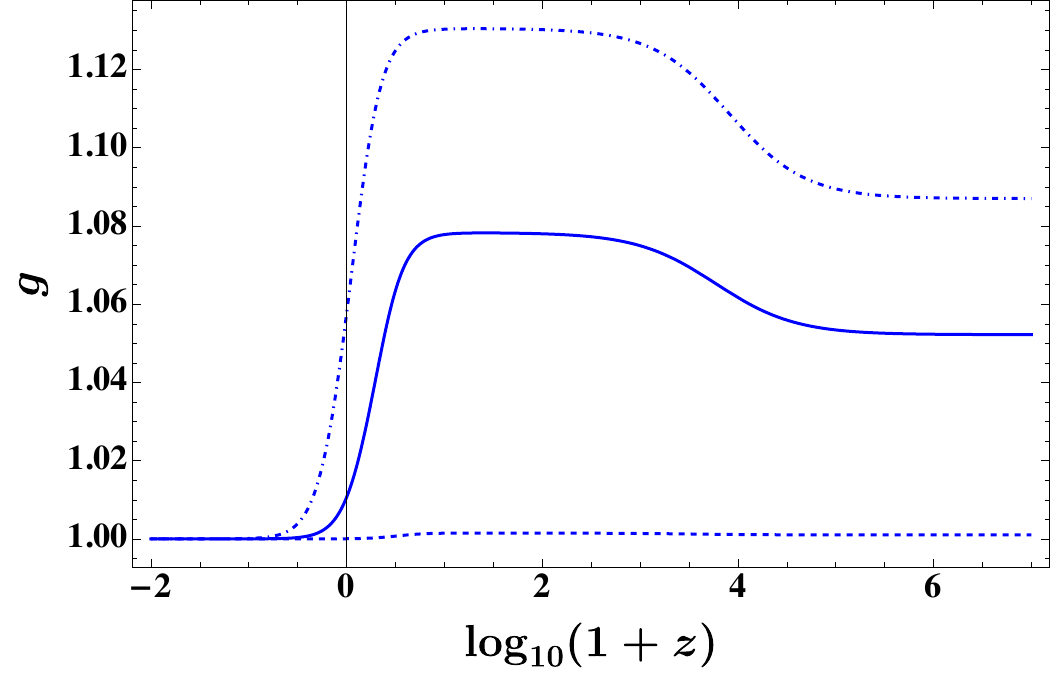} 
    \includegraphics[width=0.5\linewidth]{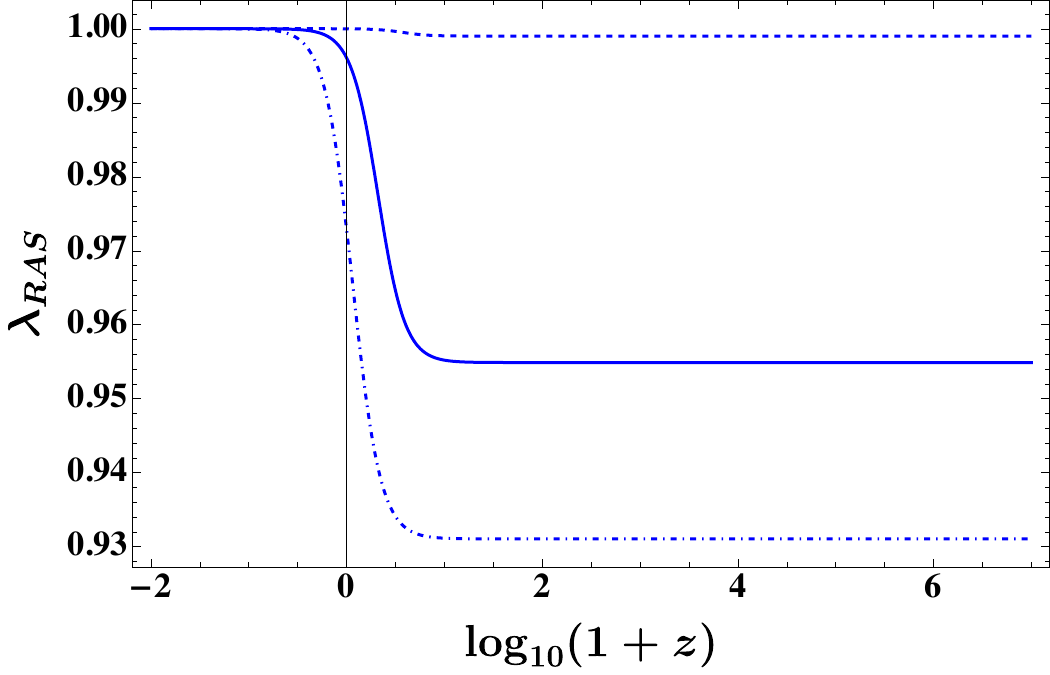} 
    \caption{\scriptsize{
    The upper panel shows the evolution of the function $g$, defined by Eq.(\ref{GG1}), as a function of $\log_{10}(1+z)$.  The lower panel shows the evolution of the Rastall parameter given by Eq.(\ref{Ras1}) as a function of the  $\log_{10}(1+z)$. As before, all panels are computed for $\alpha = -1$, considering the mean values (solid line), inferior marginalized values (dashed line) and superior marginalized values (dot-dashed line) of Table \ref{mean_inter1}, at the 68\% confidence level for \textbf{CC + PantheonPlus  + DESI}.
    }
    } 
    \label{Fig2}
\end{figure}


\begin{table*}[ht]
 \centering
 \caption{Critical points for the autonomous system for the function $f(t) = \beta\rho_m(t)$}.
\begin{center}
\begin{tabular}{c c c}\hline\hline
Name &  $\Omega_{{de}_c}$ & $\Omega_{{m}_c}$ \\\hline
$\ \ \ \ \ \ \ \ a_R \ \ \ \ \ \ \ \ $ & $0$ & $0$ \\
$\ \ \ \ \ \ \ \ b_M \ \ \ \ \ \ \ \ $ & $\frac{(\alpha +1) \beta  \left(\beta  (2 \alpha  \beta +1)+\left(2 \alpha  \beta ^2-(4 \alpha +5) \beta +1\right) w_{de}\right)}{(\alpha  \beta -1) \left(\beta +(\beta +1) w_{de}\right) \left(((4 \alpha +3) \beta -1) w_{de}-\beta \right)}$ & $\frac{-\alpha  \beta ^2 (2 \alpha  \beta +1)+\left(-2 \alpha ^2 \beta ^3+\alpha  (4 \alpha +7) \beta ^2-5 (\alpha +1) \beta +1\right) w_{de}^2+\beta  \left(-4 \alpha ^2 (\beta -1) \beta +\alpha  (6 \beta -1)+1\right) w_{de}}{(\alpha  \beta -1) \left(\beta +(\beta +1) w_{de}\right) \left(((4 \alpha +3) \beta -1) w_{de}-\beta \right)}$ \\
$\ \ \ \ \ \ \ \ c \ \ \ \ \ \ \ \ $ & $1$ & $0$ \\

\\ \hline\hline
\end{tabular}
\end{center}
\label{table1}
\end{table*}
\begin{table}[ht]
 \centering
 \caption{Cosmological parameters for the critical points in Table \ref{table1}.}
\begin{center}
\begin{tabular}{c c c c c c}\hline\hline
Name &    $\overline{\Omega}_{de}$ &    $\overline{\Omega}_{m}$ &    $\overline{\Omega}_{r}$ & $\omega_{de}$ & $\omega_{tot}$ \\\hline
$a_R$ & $0$ &  $0$ &  $1$ & const. & $1/3$ \\
$b_M$ & $\frac{(\alpha +1) \beta }{\beta -\alpha  \beta  w_{de}+w_{de}}$ &  $\frac{\alpha  \beta +(\alpha  \beta -1) w_{de}}{(\alpha  \beta -1) w_{de}-\beta }$ &  $0$ & const. & $\frac{(\alpha +1) \beta  w_{de}}{\beta -\alpha  \beta  w_{de}+w_{de}}$ \\
$c$ & $1$ &  $0$ &  $0$ & const. & $w_{de}$ \\
\\ \hline\hline
\end{tabular}
\end{center}
\label{table2}
\end{table}

\begin{itemize}
\item Critical point $a_R$  has the eigenvalues
    \begin{equation}
        \mu_1 = 1-3 w_{de}, \ \ \ \ \ \mu_2 = \frac{4 \alpha  \beta -1}{\alpha  \beta -1},
    \end{equation}
    therefore, it is unstable for
    \begin{equation}
        w_{de}<\frac{1}{3}\lor \left(\alpha \in \mathbb{R}\land \left(\left(\beta <0\land \left(\alpha <\frac{1}{\beta }\lor \alpha >\frac{1}{4 \beta }\right)\right)\lor \beta =0\lor \left(\beta >0\land \left(\alpha <\frac{1}{4 \beta }\lor \alpha >\frac{1}{\beta }\right)\right)\right)\right) .
    \end{equation}
    \item Critical point $b_M$  has the eigenvalues
    \begin{equation}
        \mu_1 = \frac{1-4 \alpha  \beta }{\alpha  \beta -1}, \ \ \ \ \ \mu_2 = -\frac{3 \alpha  \beta }{\alpha  \beta -1}-3 w_{de},
    \end{equation}
    therefore, it is unstable for
   \begin{align}
&\bigl(\beta < 0 \land \tfrac{1}{\beta} < \alpha < \tfrac{1}{4\beta}\bigr) 
 \lor \bigl(\beta > 0 \land \tfrac{1}{4\beta} < \alpha < \tfrac{1}{\beta}\bigr) \nonumber \\
&\lor \bigl(\alpha \in \mathbb{R} \land ( w_{de} < -1 \land ((\beta < 0 \land (\alpha < \tfrac{w_{de}}{\beta+\beta w_{de}} 
 \lor \alpha > \tfrac{1}{\beta})) \lor \beta = 0 \lor (\beta > 0 \land (\alpha < \tfrac{1}{\beta} 
 \lor \alpha > \tfrac{w_{de}}{\beta+\beta w_{de}}))))\bigr) \nonumber \\
&\lor (w_{de} = -1 \land ((\beta < 0 \land \alpha > \tfrac{1}{\beta}) \lor \beta = 0 
 \lor (\beta > 0 \land \alpha < \tfrac{1}{\beta}))) \nonumber \\
&\lor \bigl((-1 < w_{de} < 0 \land ((\beta < 0 \land \tfrac{1}{\beta} < \alpha < \tfrac{w_{de}}{\beta+\beta w_{de}}) 
 \lor \beta = 0 \lor (\beta > 0 \land \tfrac{w_{de}}{\beta+\beta w_{de}} < \alpha < \tfrac{1}{\beta})))) 
 \lor (w_{de} \geq 0 \land ((\beta < 0  \nonumber \\
 &\land \tfrac{1}{\beta} < \alpha < \tfrac{w_{de}}{\beta+\beta w_{de}})\lor (\beta > 0 \land \tfrac{w_{de}}{\beta+\beta w_{de}} < \alpha < \tfrac{1}{\beta}))))\bigr). \nonumber
\end{align}

    \item Critical point $c$ has the eigenvalues   
    \begin{equation}
        \mu_1 = 3 w_{de}-1, \ \ \ \ \ \mu_2 = 3 \left(\frac{\alpha  \beta }{\alpha  \beta -1}+w_{de}\right),
    \end{equation}
    therefore, it is stable for
    \begin{align}
&\bigl(\beta <0 \land \bigl((w_{de}<-1 \land (\alpha <\tfrac{w_{de}}{\beta+\beta w_{de}} \lor \alpha >\tfrac{1}{\beta})) 
 \lor (w_{de}=-1 \land \alpha >\tfrac{1}{\beta}) \lor (-1<w_{de}<\tfrac{1}{3} \land \tfrac{1}{\beta}<\alpha <\tfrac{w_{de}}{\beta+\beta w_{de}})\bigr)\bigr) \nonumber \\
&\lor \bigl(\beta =0 \land w_{de}<0\bigr) 
 \lor \bigl(\beta >0 \land ((w_{de}<-1 \land (\alpha <\tfrac{1}{\beta} \lor \alpha >\tfrac{w_{de}}{\beta+\beta w_{de}})) 
 \lor (w_{de}=-1 \land \alpha <\tfrac{1}{\beta}) \nonumber \\
 &\lor (-1<w_{de}<\tfrac{1}{3} \land \tfrac{w_{de}}{\beta+\beta w_{de}}<\alpha <\tfrac{1}{\beta})))\bigr). \nonumber
\end{align}

\end{itemize}

Figure \ref{Fig1} depicts the evolution of the system in the phase space defined by the dynamical variables $\Omega_m$ and $\Omega_{de}$ for the specific case with parameters given by the mean values of Table \ref{mean_inter1}, employing the data combination of Cosmic Chronometers (CC), type Ia Supernovae (SNe) and Baryon Acoustic Oscillations (DESI DR2). In this figure, we have utilized the value of the parameter $\alpha=-1$. The streamlines in the phase space illustrate the trajectories $a_R \to b_M \to c$, capturing the dynamical evolution of the system.

From the stability analysis of the critical points in the autonomous system, including the first interaction function $f = \beta \rho_m$, it is evident that the trajectories converge toward the attractor $c$. This convergence is clearly visible in the panel of Figure \ref{Fig1}, confirming that the corresponding critical points represent unstable ($a_R$ and $b_M$) and stable ($c$) solutions. The stable point $c$ describes a dark energy-dominated era, indicating that the first interaction model naturally guides the cosmic evolution toward a late-time attractor. In summary, the figure highlights how this interaction model within the framework of generalized Rastall gravity naturally accounts for the universe’s accelerated expansion.


The upper panel of Fig.\ref{Fig2} illustrates the evolution of the function $g(z)$ versus the variable $\log_{10}(1+z)$, for the cases where the parameters take the values of mean values (solid line), the inferior marginalized values (dashed line) and the superior marginalized values (dot-dashed line) of Table \ref{mean_inter1}, at the 68\% confidence level for CC + PantheonPlus  + DESI. From this panel, we note that the parameter $g$, as a function of the redshift, remains nearly constant and close to unity for inferior marginalized values. We also note that the function $g$ approaches the GR value, $g = 1$, in the future epoch, regardless of the value of the parameter $\beta$ used.

The lower panel shows the evolution of the Rastall parameter as a function of redshift for the same three values of the parameter $\beta$ used in the previous case. From this panel, it can be observed that the Rastall parameter $\lambda_\text{Ras}$ remains approximately constant and close to unity throughout the redshift for inferior marginalized values, while a deviation from this behavior is observed for the values of mean values and the superior marginalized values, similar to the trend shown in the central panel for the parameter $g$. Also, we observe that the Rastall parameter $\lambda_\text{Ras}(z)$ approaches its GR limit, $\lambda_\text{Ras} = 1$, in the future epoch, independently of the chosen value of the parameter $\beta$. In addition, from these panels, we observe that for small values of the parameter $\beta < 10^{-4}$ (a very small interaction), the values of the functions $g(z)$ and $\lambda_\text{Ras}(z)$ are very close to unity, and therefore, very similar to the predictions of GR.

In relation to the marginalized constraints on the parameters, we find that comparing the CC+PantheonPlus+DESI and CC+PantheonPlus data, the mean  value  Hubble parameter $H_0$ presents a small increase when we include DESI data. The same situation occurs for the parameter $\beta$ and moreover we obtain that  mean value  is positive. For the other two parameters $\Omega_m$ and $w_{de}$, we find that the inclusion of the DESI data decreases the mean values of these parameters. In particular, we note that the EoS parameter associated with the dark energy $w_{de}<-1$ (phantom behavior).

\subsection{Interaction function 
$f(t) = \beta \rho_{de}(t)$. Critical points and Stability of critical points.}

In this section, we describe the dynamical system corresponding to the second interaction, where the interaction term is defined from the function $f(t) \propto  \rho_{de}(t)$. For this case, the system can be expressed in a general form by introducing the following dimensionless variables:
 
 \begin{eqnarray}
\dfrac{d \Omega _{de}}{d N} &=& \dfrac{1}{(\alpha  \beta +\beta +1) \left(\Omega _{de} \left(2 \beta +3 w_{de}-1\right)-\Omega _m\right)} f_3 (\Omega _{de}, \Omega _{m}), 
\label{dinsyseq1b}\\
\dfrac{d \Omega _{m}}{d N} &=& \dfrac{1}{(\alpha  \beta +\beta +1) \left(\Omega _{de} \left(2 \beta +3 w_{de}-1\right)-\Omega _m\right)} f_4 (\Omega _{de}, \Omega _{m}), 
\label{dinsyseq2b}
\end{eqnarray}
where the functions $f_3(\Omega _{de}, \Omega _{m})$ and $f_4(\Omega _{de}, \Omega _{m})$ are defined as
\begin{eqnarray}
f_3 (\Omega _{de}, \Omega _{m}) &=& \Omega _{de} (\Omega _{de} (-6 w_{de} \left(\beta  \left(\alpha  \left(\Omega _m-2\right)+\Omega _m-1\right)+\Omega _m-1\right)-9 w_{de}^2+4 (\alpha +1) \beta ^2 \left(\Omega _m+2\right)+2 \beta  (\alpha  \left(\Omega _m-2\right) \nonumber\\
&&+3 \Omega _m-1)+2 \Omega _m-1)+\Omega _m \left(3 w_{de}+(\alpha +1) \beta  \left(\Omega _m-4\right)+\Omega _m-1\right)\nonumber\\
&&+(\alpha  \beta +\beta +1) \left(3 w_{de}-1\right) \Omega _{de}^2 \left(-4 \beta +3 w_{de}-1\right)), 
\label{ff3}\nonumber\\
&&\\
f_4 (\Omega _{de}, \Omega _{m}) &=&\alpha \, \beta \, \Bigl(\Omega_{de}^2 \bigl(-6 \beta -6 w_{de} (\beta +2 \beta \Omega_m+\Omega_m+1) 
   +9 w_{de}^2 (\Omega_m-1) +4 \beta \Omega_m+\Omega_m+3\bigr) \nonumber \\
&& + 2 \Omega_{de} \Omega_m \bigl(\beta -3 w_{de} (\Omega_m-1)+2 \beta \Omega_m+\Omega_m+1\bigr) 
   + (\Omega_m-1)\Omega_m^2 \Bigr) \nonumber \\
&&+ (\beta+1)\Omega_m \Bigl(\Omega_{de}\bigl(2 \beta + w_{de}(3-6 \Omega_m)+4 \beta \Omega_m+2 \Omega_m-1\bigr) 
   + (3 w_{de}-1)\Omega_{de}^2(-4 \beta+3 w_{de}-1) \nonumber \\
&& + (\Omega_m-1)\Omega_m \Bigr), 
\label{ff4}
\end{eqnarray}
respectively.

Additionally, from Eq.(\ref{GA}), we obtain that the function $g(\Omega _{de}, \Omega _{m})$, expressed in terms of the dynamical variables for this interaction model, takes the following form:

\begin{eqnarray}
   g(\Omega _{de}, \Omega _{m})= g =  \frac{\left(1-3 w_{de}\right) \Omega _{de}+\Omega _m}{\Omega _m-\Omega _{de} \left(2 \beta +3 w_{de}+\beta  \Omega _m-1\right)+\beta  \left(3 w_{de}-1\right) \Omega _{de}^2}.\label{GG2}
\end{eqnarray}

Moreover, we find that for our interaction model, the Rastall parameter derived from Eq.(\ref{Ras}) takes the form
\begin{equation}
\lambda_\text{Ras}(\Omega_{de},\Omega_m)=\lambda_\text{Ras}=\left[\frac{2\beta\Omega_{de}+(1-3w_{de})\Omega_{de}+\Omega_m}{4\beta\Omega_{de}+(1-3w_{de})\Omega_{de}+\Omega_m}\right].\label{Ras2}
\end{equation}

On the other hand, 
based on the functions $f_3(\Omega_{de}, \Omega_m)$ and $f_4(\Omega_{de}, \Omega_m)$, the critical points corresponding to the second interaction model are obtained, by setting the derivatives $d\Omega_{de}/dN$ and $d\Omega_{m}/dN$ equal to zero in Eqs.(\ref{dinsyseq1b}) and (\ref{dinsyseq2b}), respectively. As before, these conditions allow us to identify the stationary solutions that characterize the dynamical behavior of the system.
In this sense, Table \ref{table3} lists the critical points of the dynamical system corresponding to the second interaction function, $f = \beta \rho_{de}$, within the framework of generalized Rastall gravity. Table \ref{table4} presents the associated values of the cosmological parameters for each of these points. As before, the quantities $\bar{\Omega}_{de}$ and $\bar{\Omega}_m$ are defined according to Eq.(\ref{bO}).

 In these tables, $d_R$ denotes the critical point associated with a radiation-dominated epoch. As in the previous case, at this point, the radiation density parameter satisfies $\bar{\Omega}_r = 1$, and the total EoS parameter is $\omega_{tot} = 1/3$. This critical point is independent of both; the value of $\alpha$ and the interaction parameter $\beta$ associated with the function $f$.

In these tables, the critical point $f_M$ corresponds to a matter-dominated epoch, characterized by $\bar{\Omega}_m = 1$ and a total EoS parameter $\omega_{tot} = 0$. Similarly to the radiation-dominated point, this critical point does not depend on the values of $\alpha$ or the interaction parameter $\beta$ related to the function $f$.

The critical point $h$ represents a dark energy era, where the total EoS parameter depends on the interaction parameters $\alpha$ and $\beta$. In particular, for a non-interacting model where the parameter $\beta=0$, the cosmological parameter $\bar{\Omega}_{de}=1$ and $\omega_{tot}=w_{de}$. However, the case $\beta = 0$ represents a scenario without any interaction and is thus excluded from the present analysis. In addition, for the particular case in which $w_{de}=-1$, we also find that the cosmological parameter $\bar{\Omega}_{de}=1$ and the total EoS parameter $\omega_{tot}=-1$ corresponding to a de Sitter solution.

In what follows, we will present the eigenvalues associated with the critical points of the second interaction function model within the framework of generalized Rastall gravity with the conditions that determine their stability.

\begin{table*}[ht]
 \centering
 \caption{Critical points for the autonomous system for $f(t) = \beta \rho_{de}(t)$}.
\begin{center}
\begin{tabular}{c c c}\hline\hline
Name &  $\Omega_{de}$ & $\Omega_{m}$ \\\hline
$\ \ \ \ \ \ \ \ d_R \ \ \ \ \ \ \ \ $ & $0$ & $0$ \\
$\ \ \ \ \ \ \ \ f_M \ \ \ \ \ \ \ \ $ & $0$ & $1$ \\
$\ \ \ \ \ \ \ \ h \ \ \ \ \ \ \ \ $ & $\frac{(\alpha +1) \beta ^2 (2 (\alpha +1) \beta -1)+\beta  \left(4 \alpha ^2 \beta +3 \alpha  \beta +\alpha -\beta +2\right) w_{de}-((7 \alpha +4) \beta +1) w_{de}^2+3 w_{de}^3}{(\alpha  \beta +\beta +1) \left(w_{de}-\beta \right) \left(\beta -((4 \alpha +3) \beta +1) w_{de}+3 w_{de}^2\right)}$ & $\frac{\alpha  \beta  \left(w_{de}+1\right) \left(-2 (\alpha +1) \beta ^2+\beta -(4 \alpha  \beta +\beta +1) w_{de}+3 w_{de}^2\right)}{(\alpha  \beta +\beta +1) \left(w_{de}-\beta \right) \left(\beta -((4 \alpha +3) \beta +1) w_{de}+3 w_{de}^2\right)}$ \\

\\ \hline\hline
\end{tabular}
\end{center}
\label{table3}
\end{table*}
\begin{table}[ht]
 \centering
 \caption{Cosmological parameters for the critical points in Table \ref{table3} in which $f\propto \rho_{de}$.}
\begin{center}
\begin{tabular}{c c c c c c}\hline\hline
Name &    $\overline{\Omega}_{de}$ &    $\overline{\Omega}_{m}$ &    $\overline{\Omega}_{r}$ & $\omega_{de}$ & $\omega_{tot}$ \\\hline
$d_R$ & $0$ &  $0$ &  $1$ & const. & $1/3$ \\
$f_M$ & $0$ &  $1$ &  $0$ & const. & $0$ \\
$h$ & $\frac{w_{de}-(\alpha +1) \beta }{(\alpha  \beta +1) w_{de}-\beta }$ &  $\frac{\alpha  \beta  \left(w_{de}+1\right)}{(\alpha  \beta +1) w_{de}-\beta }$ &  $0$ & const. & $\frac{w_{de} \left(w_{de}-(\alpha +1) \beta \right)}{(\alpha  \beta +1) w_{de}-\beta }$ \\
\\ \hline\hline
\end{tabular}
\end{center}
\label{table4}
\end{table}

\begin{figure}[!h]
  \centering
    \includegraphics[width=0.5\linewidth]{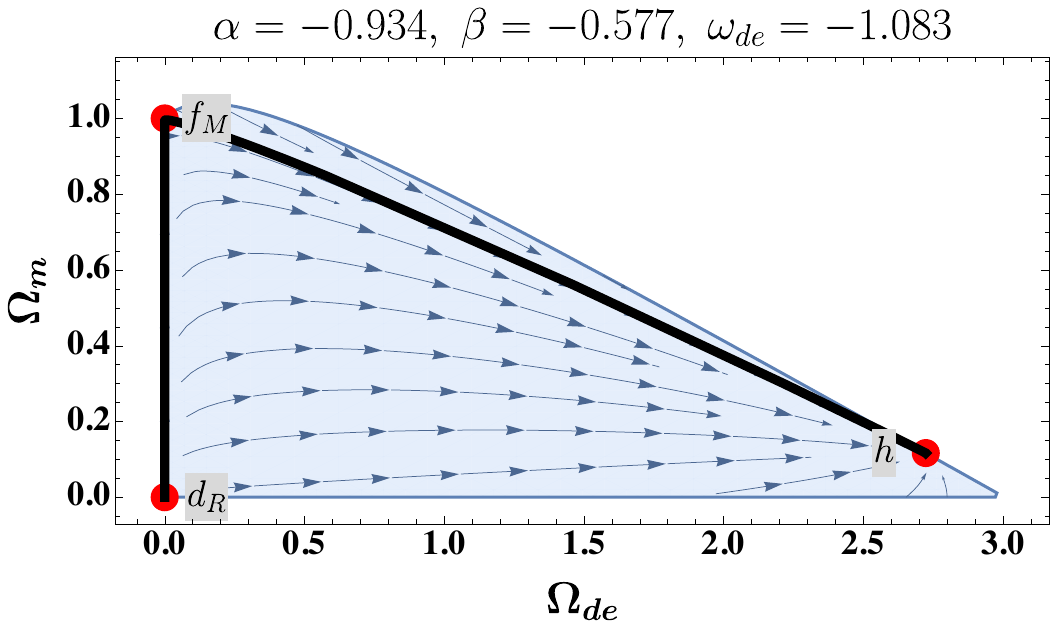} 
    \caption{\scriptsize{This figure depicts the phase-space evolution $(\Omega_m,\Omega_{de})$ for the  second interaction function $f\propto\rho_{de}$ within generalized Rastall gravity, with parameters given by the mean values of Table \ref{mean_inter2} for \textbf{CC + PantheonPlus  + DESI}.
    }
    } 
    \label{Fig3}
\end{figure}

\begin{figure}[!h]
  \centering
    \includegraphics[width=0.5\linewidth]{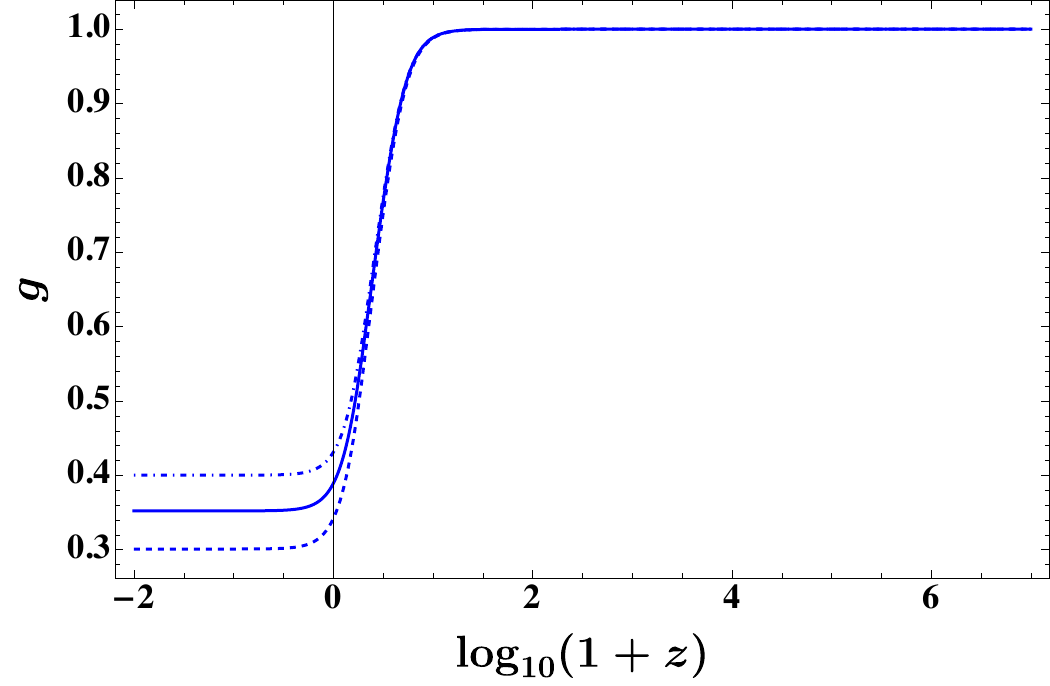} 
     \includegraphics[width=0.5\linewidth]{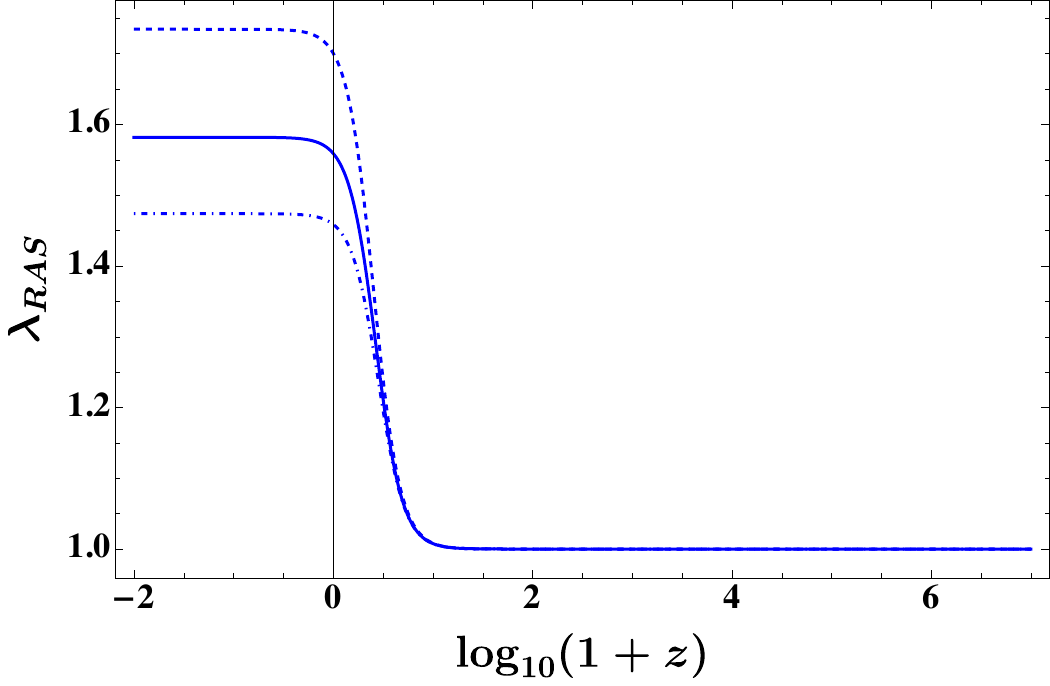} 
    \caption{\scriptsize{As before, the upper panel 
    shows the evolution of the function $g$, defined by Eq.(\ref{GG2}), as a function of $\log_{10}(1+z)$.  The lower panel displays  the evolution of the Rastall parameter given by Eq.(\ref{Ras2}) as a function of the  $\log_{10}(1+z)$. In all panels we have used the mean values (solid line), inferior marginalized values (dashed line) and superior marginalized values (dot-dashed line) of Table \ref{mean_inter2}, at the 68\% confidence level for \textbf{CC + PantheonPlus  + DESI}.   }
    } 
    \label{Fig4}
\end{figure}

\begin{itemize}
    \item Point $d_R$ has the eigenvalues   
    \begin{equation}
        \mu_1 = 1, \ \ \ \ \ \mu_2 =-\frac{-4 (\alpha +1) \beta +3 w_{de}-1}{\alpha  \beta +\beta +1},
    \end{equation}
    therefore, it is unstable.
    \item Point $f_M$  has the eigenvalues
    \begin{equation}
        \mu_1 = -1, \ \ \ \ \ \mu_2 = \frac{3 \left(\alpha  \beta +\beta -w_{de}\right)}{\alpha  \beta +\beta +1},
    \end{equation}
    therefore, it is unstable for
   \begin{align}
&\bigl(w_{de}\leq -1 \land \bigl((\alpha<-1 \land (\beta<-\tfrac{1}{\alpha+1} \lor \beta>\tfrac{w_{de}}{\alpha+1})) 
   \lor \alpha=-1 \lor (\alpha>-1 \land (\beta<\tfrac{w_{de}}{\alpha+1} \lor \beta>-\tfrac{1}{\alpha+1})))\bigr) \nonumber \\
&\lor \bigl(-1<w_{de}<0 \land \bigl((\alpha<-1 \land (\beta<\tfrac{w_{de}}{\alpha+1} \lor \beta>-\tfrac{1}{\alpha+1})) 
   \lor \alpha=-1 \lor (\alpha>-1 \land (\beta<-\tfrac{1}{\alpha+1} \lor \beta>\tfrac{w_{de}}{\alpha+1})))\bigr) \nonumber \\
&\lor \bigl(w_{de}\geq 0 \land \bigl((\alpha<-1 \land (\beta<\tfrac{w_{de}}{\alpha+1} \lor \beta>-\tfrac{1}{\alpha+1})) 
   \lor (\alpha>-1 \land (\beta<-\tfrac{1}{\alpha+1} \lor \beta>\tfrac{w_{de}}{\alpha+1})))\bigr) \nonumber
\end{align}

    \item Point $h$ has the eigenvalues   
    \begin{equation}
        \mu_1 = \frac{-4 (\alpha +1) \beta +3 w_{de}-1}{\alpha  \beta +\beta +1}, \ \ \ \ \ \mu_2 = \frac{3 \left(w_{de}-(\alpha +1) \beta \right)}{\alpha  \beta +\beta +1},\label{u12}
    \end{equation}
    therefore, it is stable for
    \begin{align}
&\bigl(w_{de} \leq -1 \land ((\alpha < -1 \land (\beta < -\tfrac{1}{1+\alpha} \lor \beta > \tfrac{w_{de}}{1+\alpha})) 
   \lor \alpha = -1 \lor (\alpha > -1 \land (\beta < \tfrac{w_{de}}{1+\alpha} \lor \beta > -\tfrac{1}{1+\alpha})))\bigr) \nonumber \\
&\lor \bigl(-1 < w_{de} < 0 \land ((\alpha < -1 \land (\beta < \tfrac{w_{de}}{1+\alpha} \lor \beta > -\tfrac{1}{1+\alpha})) 
   \lor \alpha = -1 \lor (\alpha > -1 \land (\beta < -\tfrac{1}{1+\alpha} \lor \beta > \tfrac{w_{de}}{1+\alpha})))\bigr) \nonumber \\
&\lor \bigl(w_{de} \geq 0 \land ((\alpha < -1 \land (\beta < \tfrac{w_{de}}{1+\alpha} \lor \beta > -\tfrac{1}{1+\alpha})) 
   \lor (\alpha > -1 \land (\beta < -\tfrac{1}{1+\alpha} \lor \beta > \tfrac{w_{de}}{1+\alpha}))))\bigr) .\nonumber
\end{align}
\end{itemize}

Figure \ref{Fig3} illustrates the phase-space evolution of the system in terms of the dynamical variables $\Omega_{m}$ and $\Omega_{de}$. The plot corresponds to phase-space  for the interaction function $f\propto\rho_{de}$
together with parameters given by the mean values of Table \ref{mean_inter2}, employing the data CC, SNe and DESI DR2.
As before, we show 
that the phase space is characterized by  the trajectories $d_R \rightarrow f_M \rightarrow h$, which describe  the dynamical evolution of the system.

The stability analysis of the autonomous system associated  to the second interaction function $f=\beta\rho_{de}$ indicates  that all trajectories in the phase space develop  toward the attractor point $h$ (dark energy epoch). This behavior is shown in  Fig.(\ref{Fig3}), indicating  that the critical points
$d_R$ and $f_M$ are to unstable configurations, while the point critical $h$ corresponds to a stable attractor solution. Thus, from this figure, the stability analysis of the critical points associated with  the second interaction reveals that the system approaches a late-time attractor and corresponds  to a dark energy dominated epoch, in which  the cosmological parameter $\bar{\Omega}_{de}=1$ at the current epoch.

On the other hand, the upper and lower panels of Fig.\ref{Fig4} show  the evolution of the function $g(z)$ and  the Rastall parameter $\lambda_\text{Ras}(z)$ versus $\log_{10}(1+z)$. In this figure we have used  three  representative specific values of the parameters corresponding as before to the values of mean values (solid line), the inferior marginalized values (dashed line) and the superior marginalized values (dot-dashed line) of Table \ref{mean_inter2}, at the 68\% confidence level (at 1-$\sigma$) for CC + PantheonPlus  + DESI.

From the upper and lower panels of Fig.\ref{Fig4},  we observe  that the parameters $g(z)$ and $\lambda_\text{Ras}(z)$, as a function of the redshift, remain nearly constant and close to unity, when the redshift $z>10$. This result becomes independent of the values of the parameters of our model. However, for values of the redshift $z<9$ and in particular at the present epoch, we note that the interaction-model $f\propto \rho_{de}$ deviates from GR using the best values of our  parameters   at 1-$\sigma$ assuming  CC + PantheonPlus  + DESI.  

Regarding to the marginalized constraints on the parameters for our second interaction, we obtain that comparing the CC+PantheonPlus+DESI and CC+PantheonPlus data, the mean  value  Hubble parameter $H_0$ presents a small increase when we include DESI data. The same situation takes place for the parameter $\beta$ and moreover we find that  mean value  is negative and around the value of $-0.6$. For the other two parameters $\Omega_m$ and $w_{de}$, we find that the inclusion of the DESI data increases the mean values of these parameters. Also, as before, we note that the EoS parameter is associated with dark energy $w_{de}<-1$ (phantom behavior). In relation to the parameter $\alpha$, we find that the mean value obtained is $-1.037$ using CC+PantheonPlus and $-0.935$ by adding DESI data (see Table \ref{mean_inter2}).

\subsection{Interaction function $f(t) = \beta [\,\rho_m(t)+\rho_{de}(t)\,]$. Critical points and Stability of critical points.}

In this section, we study the dynamical system associated  with  our third interaction function, given by $f=\beta\,(\rho_m+\rho_{de})$ with $\beta$ a constant parameter. From this interaction and $\alpha=-1$ (otherwise, the matter-dominated era is not recovered), the dynamical system can be written in a general form and its evolution is given by the following set of equations:
 \begin{eqnarray}
\dfrac{d \Omega _{de}}{d N} &=& \dfrac{\Omega _{de}}{\Omega _{de} \left(2 \beta +3 w_{de}-1\right)+(2 \beta -1) \Omega _m} f_5 (\Omega _{de}, \Omega _{m}), 
\label{dinsyseq1b}\\
\dfrac{d \Omega _{m}}{d N} &=& \dfrac{1}{(\beta +1) \left(\Omega _{de} \left(2 \beta +3 w_{de}-1\right)+(2 \beta -1) \Omega _m\right)} f_6 (\Omega _{de}, \Omega _{m}), 
\label{dinsyseq2b}
\end{eqnarray}
where the functions $f_5(\Omega _{de}, \Omega _{m})$ and $f_6(\Omega _{de}, \Omega _{m})$ become
\begin{eqnarray}
f_5 (\Omega _{de}, \Omega _{m}) &=&  \Omega _{de} \left(2 \beta -6 w_{de} \left(\beta +2 \beta  \Omega _m+\Omega _m-1\right)-9 w_{de}^2+8 \beta  \Omega _m+2 \Omega _m-1\right)\nonumber\\
&& +\Omega _m \left(2 \beta +(3-6 \beta ) w_{de}+4 \beta  \Omega _m+\Omega _m-1\right)+\left(3 w_{de}-1\right) \Omega _{de}^2 \left(-4 \beta +3 w_{de}-1\right),\nonumber\\
&&\\
f_6 (\Omega _{de}, \Omega _{m}) &=& -2 \beta ^2 \left(\Omega _{de}+\Omega _m\right) \left(\Omega _{de} \left(6 w_{de} \Omega _m-3 w_{de}-2 \Omega _m-3\right)-2 \Omega _m \left(\Omega _m+2\right)\right) \nonumber \\
&& +\beta  \left(\left(3 w_{de}-1\right) \Omega _{de}^2 \left(3 w_{de} \left(\Omega _m+1\right)-5 \Omega _m+3\right)-\left(9 w_{de}-5\right) \Omega _{de} \Omega _m \left(2 \Omega _m-1\right)+\Omega _m^2 \left(5 \Omega _m-2\right)\right)\nonumber \\
&& +\Omega _m \left(\Omega _{de} \left(-6 w_{de} \Omega _m+3 w_{de}+2 \Omega _m-1\right)+\left(1-3 w_{de}\right){}^2 \Omega _{de}^2+\left(\Omega _m-1\right) \Omega _m\right),\nonumber \\
&&
\label{ff4}
\end{eqnarray}
respectively.

From this interaction and considering   Eq.(\ref{GA}), we find that the function $g(\Omega _{de}, \Omega _{m})$, expressed in terms of the dynamical variables results

\begin{eqnarray}
   g(\Omega _{de}, \Omega _{m})= g = \frac{\left(1-3 w_{de}\right) \Omega _{de}+\Omega _m}{-2 \beta  \Omega _{de} \left(\Omega _m+1\right)+3 w_{de} \Omega _{de} \left(\beta  \Omega _m-1\right)+\beta  \left(3 w_{de}-1\right) \Omega _{de}^2+\Omega _{de}-\beta  \Omega _m \left(\Omega _m+2\right)+\Omega _m} .\label{GG3}
\end{eqnarray}

Moreover, we find that for our interaction model, the Rastall parameter derived from Eq.(\ref{Ras}) is defined as
\begin{equation}
\lambda_\text{Ras}(\Omega_{de},\Omega_m)=\lambda_\text{Ras}= \frac{\Omega _{de} \left(2 \beta -3 w_{de}+1\right)+(2 \beta +1) \Omega _m}{\Omega _{de} \left(4 \beta -3 w_{de}+1\right)+(4 \beta +1) \Omega _m}.
\label{Ras3}
\end{equation}

As before, using this dynamical system, we will find the critical points associated with our third interaction, under the conditions $d\Omega_m/dN=d\Omega_{de}/dN=0$. In this way, Table \ref{table5} shows the different critical points found for our third interaction function $f\propto \rho_m+\rho_{de}$. In addition, Table \ref{table6} presents the corresponding values of their cosmological parameters; $\bar{\Omega}_{de}$, $\bar{\Omega}_m$, $\bar{\Omega}_r$, $w_{de}$ and $\omega_{tot}$ in relation to the critical points.

In these tables, the critical point $i_R$ denotes a radiation era. In this point, the total EoS parameter $\omega_{tot}=1/3$ and the radiation parameter $\bar{\Omega}_r=1$, whereas the other two parameters $\bar{\Omega}_{de}$ and $\bar{\Omega}_m$ are zero. Additionally, we note that this point $i_R$ does not depend on the parameters $\alpha$ and $\beta$ associated with the interaction term $Q_1=\alpha\beta[\dot{\rho}_m+\dot{\rho}_{de}]$ (or $Q_2$).

The critical point $j_M$ associated with the matter-dominated  epoch is found only when the parameter $\alpha=-1$, 
as in the first interaction. For this point $j_M$, the density parameter $\bar{\Omega}_m=1$ and the total EoS parameter takes the value zero i.e., a pressureless fluid.
In this sense, to obtain a matter-dominated epoch, our interaction model fixes the value of the parameter $\alpha$ to minus one. As before, this condition implies that the model features an interaction only with dark matter, where the interaction term $Q_2 = 0$, (see Eqs.(\ref{Q2})). In this case, the interaction takes place only in the matter equation of motion, and not in the dark energy component. In this context, as we mentioned previously,  this type of interaction 
can be related with particle production, see e.g. Refs.\cite{Prigogine:1989zz,NunesPavon2015_PRD_ParticleCreation,Parker:1968mv,Parker:1971pt,Calvao:1991wg,Pramanik:2025jwe}.

The other  critical point corresponds to the point $k$ and represents a dark energy–dominated era, where the cosmological parameter  $\bar{\Omega}_{de}$ depends on the parameter $w_{de}$ and the coupling parameter $\beta$. The critical point $k$ corresponds to a stable point with eigenvalues $\mu_1$ and $\mu_2$ defined by Eq.(\ref{u12}). In particular, for the specific case in which the EoS parameter $w_{de}=-1$ (cosmological constant), we have the cosmological parameter associated with matter $\bar{\Omega}_m=0$ and $\bar{\Omega}_{de}=1$.

\begin{table*}[ht]
 \centering
 \caption{Critical points for the autonomous system for $f(t) = \beta [\,\rho_m(t)+\rho_{de}(t)\,]$}.
\begin{center}
\begin{tabular}{c c c}\hline\hline
Name &  $\Omega_{de}$ & $\Omega_{m}$ \\\hline
$\ \ \ \ \ \ \ \ i_R \ \ \ \ \ \ \ \ $ & $0$ & $0$ \\
$\ \ \ \ \ \ \ \ j_M \ \ \ \ \ \ \ \ $ & $0$ & $\frac{1-2 \beta }{\beta +1}$ \\
$\ \ \ \ \ \ \ \ k \ \ \ \ \ \ \ \ $ & $\frac{\beta  (5 \beta -1)+\left(8 \beta ^2+7 \beta -1\right) w_{de}+3 (\beta +1)^2 w_{de}^2}{(\beta +1) w_{de} \left(3 \beta +3 (\beta +1) w_{de}-1\right)}$ & $-\frac{\beta  \left(w_{de}+1\right) \left(5 \beta +3 (\beta +1) w_{de}-1\right)}{(\beta +1) w_{de} \left(3 \beta +3 (\beta +1) w_{de}-1\right)}$ \\

\\ \hline\hline
\end{tabular}
\end{center}
\label{table5}
\end{table*}
\begin{table}[ht]
 \centering
 \caption{Cosmological parameters for the critical points in Table \ref{table5}.}
\begin{center}
\begin{tabular}{c c c c c c}\hline\hline
Name &    $\overline{\Omega}_{de}$ &    $\overline{\Omega}_{m}$ &    $\overline{\Omega}_{r}$ & $\omega_{de}$ & $\omega_{tot}$ \\\hline
$i_R$ & $0$ &  $0$ &  $1$ & const. & $1/3$ \\
$j_M$ & $0$ &  $1$ &  $0$ & const. & $0$ \\
$k$ & $1+\beta +\frac{\beta }{w_{de}}$ &  $-\frac{\beta  \left(w_{de}+1\right)}{w_{de}}$ &  $0$ & const. & $\beta +(\beta +1) w_{de}$ \\
\\ \hline\hline
\end{tabular}
\end{center}
\label{table6}
\end{table}

\begin{figure}[!h]
  \centering
    \includegraphics[width=0.5\linewidth]{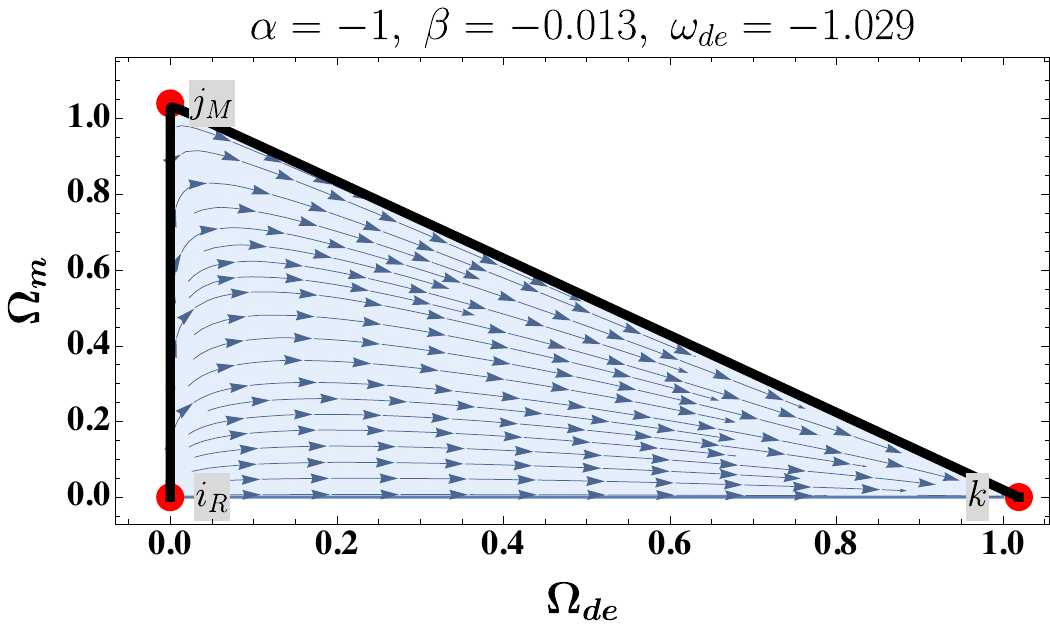} 
    \caption{\scriptsize{This figure depicts the phase-space evolution $(\Omega_m,\Omega_{de})$ for the first interaction function within generalized Rastall gravity, with parameters given by the mean values of Table \ref{mean_inter3} for \textbf{CC + PantheonPlus  + DESI}.
    }
    } 
    \label{Fig5}
\end{figure}

\begin{figure}[!h]
  \centering
    \includegraphics[width=0.5\linewidth]{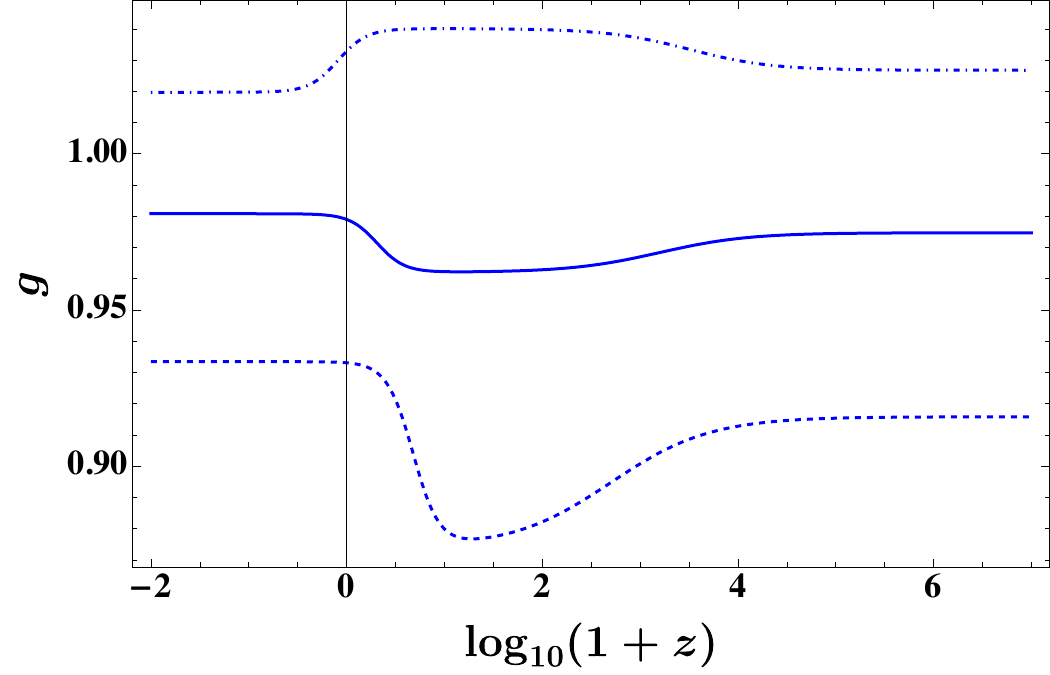} 
     \includegraphics[width=0.5\linewidth]{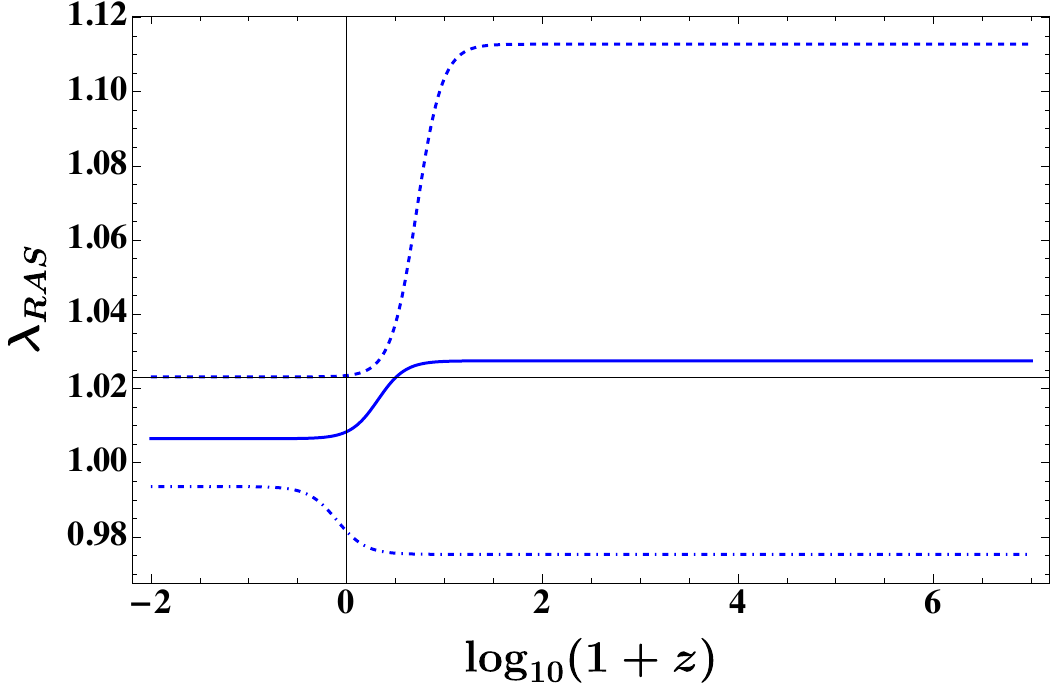} 
    \caption{\scriptsize{As before, the upper panel 
    shows the evolution of the function $g$, defined by Eq.(\ref{GG3}), as a function of $\log_{10}(1+z)$.  The lower panel displays  the evolution of the Rastall parameter given by Eq.(\ref{Ras3}) as a function of the  $\log_{10}(1+z)$. In all panels we have used the mean values (solid line), inferior marginalized values (dashed line) and superior marginalized values (dot-dashed line) of Table \ref{mean_inter2}, at the 68\% confidence level for \textbf{CC + PantheonPlus  + DESI}.   }
    } 
    \label{Fig6}
\end{figure}

\begin{itemize}
    \item Point $i_R$ has the eigenvalues   
    \begin{equation}
        \mu_1 = 1-3 w_{de}, \ \ \ \ \ \mu_2 =4-\frac{3}{\beta +1},
    \end{equation}
    therefore, it is unstable for
    \begin{equation}
        w_{de}<\frac{1}{3}\lor \beta <-1\lor \beta >-\frac{1}{4}
    \end{equation}
    \item Point $j_M$  has the eigenvalues
    \begin{equation}
        \mu_1 = -\frac{4 \beta +1}{\beta +1}, \ \ \ \ \ \mu_2 = -\frac{3 \left(\beta +\beta  w_{de}+w_{de}\right)}{\beta +1},
    \end{equation}
    therefore, it is unstable for
   \begin{align}
&-1<\beta <-\frac{1}{4}\lor \left(w_{de}<-1\land \left(\beta <-\frac{w_{de}}{w_{de}+1}\lor \beta >-1\right)\right)\lor \left(w_{de}=-1\land \beta >-1\right) \nonumber \\
&\lor \left(w_{de}>-1\land -1<\beta <-\frac{w_{de}}{w_{de}+1}\right) \nonumber
\end{align}

    \item Point $k$ has the eigenvalues   
    \begin{equation}
        \mu_1 = -1+3 w_{de}, \ \ \ \ \ \mu_2 = \frac{3 \left(\beta +\beta  w_{de}+w_{de}\right)}{\beta +1},
    \end{equation}
    therefore, it is stable for
    \begin{align}
&\left(w_{de}<-1\land \left(\beta <-\frac{w_{de}}{w_{de}+1}\lor \beta >-1\right)\right)\lor \left(w_{de}=-1\land \beta >-1\right)\lor \left(-1<w_{de}<\frac{1}{3}\land -1<\beta <-\frac{w_{de}}{w_{de}+1}\right). \nonumber \\
\end{align}
\end{itemize}

Figure \ref{Fig5} shows the evolution of the system in the phase space given by the dynamical variables $\Omega_m$ and $\Omega_{de}$.
In order to study this phase space, we have used as before
 the mean values of Table \ref{mean_inter3} at 1-$\sigma$.   In addition, in this figure, we have assumed the value of the parameter $\alpha=-1$. The flow lines in this figure represent  the different trajectories $i_R \to j_M \to k$, characterizing  the dynamical evolution of the our interaction model.

By considering the stability analysis of the critical points by  the third interaction function $f \propto  \rho_m+\rho_{de}$, we note that the flow lines  converge to the attractor $k$. From this convergence (see Fig.\ref{Fig5}), we observe  that the corresponding critical points $i_R$ and $j_M$ correspond to unstable solutions,  while $k$ is a stable solution. Here, the stable point $k$ characterizes  a dark energy-dominated epoch, so that the third  interaction model  drives the cosmic evolution toward a late-time attractor and then to an  accelerated expansion of the late-time Universe.

On the other hand, the upper panel of Fig.\ref{Fig6} shows the evolution of the function $g(z)$ in terms of the variable $\log_{10}(1+z)$.
In this panel, we have considered 
 the parameters take the values of mean values (solid line), the inferior marginalized values (dashed line) and the superior marginalized values (dot-dashed line) of Table \ref{mean_inter3}, at the 68\% confidence level  (1-$\sigma$).

Observing  this panel, we deduce  that the parameter $g$, as a function of the redshift, remains nearly constant for large values of the redshift ($z>10^4$).
 We also observe that the function $g$ does not approach the GR value, in which  $g = 1$. This suggests that for this third interaction the model does not recover  the GR limit, but rather exhibits a small difference.
 
 The lower panel of Fig.\ref{Fig6} shows the evolution of the Rastall parameter as a function of redshift for the same three values of the parameter $\beta$ used in the upper panel i.e., at 1-$\sigma$. From this panel, it can be noted that the Rastall parameter $\lambda_\text{Ras}$ remains approximately constant for redshift values $z>30$.
 
  Also, we observe that the Rastall parameter $\lambda_\text{Ras}(z)$ does not approach to GR limit, where $\lambda_\text{Ras} = 1$ for redshift $z\sim 0$. This suggests that for the interaction $f\propto \rho_m+\rho_{de}$, generalized Rastall gravity presents a slight deviation in relation to GR theory, since $\lambda_\text{Ras}(z=0)\simeq 1$ using the values of the parameters at 1-$\sigma$, from Table \ref{mean_inter3}.
  
  From the marginalized constraints on the parameters described by the Table \ref{mean_inter3}, we obtain  that comparing the CC+PantheonPlus+DESI and CC+PantheonPlus data, the mean  value  Hubble parameter $H_0$ presents a small decrease when we include DESI data. In relation to the interaction parameter $\beta$  we find that  mean value  is negative, as in the first interaction function. For the other two parameters $\Omega_m$ and $w_{de}$, we get that the inclusion of the DESI data decreases the mean values of the density parameter $\Omega_m$ and increases the value of EoS parameter $w_{de}$. In addition,  we note  that the EoS parameter associated with the dark energy $w_{de}<-1$  and it shows a phantom-like behavior.

\section{Statistical analysis: $\chi^2$ estimators}\label{ST}

In order to constrain the free parameters of the model, we perform a joint likelihood analysis based on three independent cosmological probes: Cosmic Chronometers (CC), Type Ia Supernovae (SNe), and Baryon Acoustic Oscillations (BAO) from the DESI survey. 
Each dataset provides complementary information on the cosmic expansion history and allows us to break parameter degeneracies. 
For all cases, the total chi-square $\chi^2_{\rm tot}$ is defined as
\begin{equation}
\chi^2_{\rm tot} = \chi^2_{\rm CC} + \chi^2_{\rm SN} + \chi^2_{\rm DESI}.
\end{equation}
In the following, we will define the individual chi-squared  values.

\vspace{0.3cm}
\noindent
\textbf{Cosmic Chronometers (CC):}
The CC method relies on differential age measurements of passively evolving galaxies to obtain a direct determination of the Hubble parameter $H(z)$, given $N_{\rm CC}$ observational points $\{z_i,\, H_{\rm obs}(z_i),\, \sigma_{H_i}\}$, for details see \cite{Moresco:2020fbm,cao2018cosmological,farooq2013hubble}. The corresponding chi-square estimator reads
\begin{equation}
\chi^2_{\rm CC} = \sum_{i=1}^{N_{\rm CC}}
\frac{\left[H_{\rm th}(z_i;\mathbf{p}) - H_{\rm obs}(z_i)\right]^2}{\sigma_{H_i}^2},
\end{equation}
where $H_{\rm th}(z_i;\mathbf{p})$ denotes the theoretical prediction of the model for a given set of parameters $\mathbf{p} = \{H_0,\Omega_m,\alpha,\beta,\omega_{de} \}$.

\vspace{0.3cm}
\noindent
\textbf{Type Ia Supernovae (SNe):}
Supernovae provide measurements of the luminosity distance $D_L(z)$ through the observed distance modulus
\begin{equation}
\mu_{\rm obs} = m_B^{\rm corr} - M_B,
\end{equation}
where $m_B^{\rm corr}$ is the corrected apparent magnitude and $M_B$ is the absolute magnitude of the standard candle. 
The theoretical counterpart is given by
\begin{equation}
\mu_{\rm th}(z_i) = 5\log_{10}\!\left[\frac{D_L(z_i;\mathbf{p})}{10\,{\rm pc}}\right] + 25,
\quad
D_L(z) = c(1+z)\int_0^z \frac{dz'}{H(z';\mathbf{p})}.
\end{equation}
Thus, the chi-square for the SNe dataset \cite{Brownsberger:2021uue,Popovic:2021yuo,Carr:2021lcj,Peterson:2021hel,Brout:2021mpj,Scolnic:2021amr,Brout:2022vxf,Yuan_2022,Riess:2021jrx}\footnote{Data available online in the GitHub repository \url{https://github.com/PantheonPlusSH0ES}.} can be expressed as
\begin{equation}
\chi^2_{\rm SN} = 
(\boldsymbol{\mu}_{\rm obs}-\boldsymbol{\mu}_{\rm th})^{\!T}
\,\mathbf{C}^{-1}\,
(\boldsymbol{\mu}_{\rm obs}-\boldsymbol{\mu}_{\rm th}),
\end{equation}
where $\mathbf{C}$ is the full covariance matrix including statistical and systematic uncertainties.
We use the PantheonPlus dataset where we fix $M_B = -19.253$, allowing $H_0$ to be directly constrained.

\vspace{0.3cm}
\noindent
\textbf{Baryon Acoustic Oscillations (DESI):}
BAO measurements from the DESI \cite{DESI:2025zgx,DESI:2025zpo}\footnote{Data available online in the GitHub repository \url{https://github.com/CobayaSampler/bao_data/}.} survey provide constraints on the comoving distances 
normalized by the comoving sound horizon at the drag epoch $r_d$, namely
\begin{equation}
\mathbf{X}_{\rm obs} =
\left\{\frac{D_M(z)}{r_d},\,\frac{D_H(z)}{r_d},\,\frac{D_V(z)}{r_d}\right\}_{\!\!{\rm obs}},
\end{equation}
where $D_M(z)$ is the comoving angular diameter distance, 
$D_H(z) = c/H(z)$ is the Hubble distance, 
and $D_V(z)$ is the volume-averaged distance.
The corresponding chi-square function related to DESI is defined as
\begin{equation}
\chi^2_{\rm DESI} =
(\mathbf{X}_{\rm obs} - \mathbf{X}_{\rm th})^{\!T}
\,\mathbf{C}^{-1}\,
(\mathbf{X}_{\rm obs} - \mathbf{X}_{\rm th}),
\quad
\mathbf{X}_{\rm th} =
\left\{
\frac{D_M(z)}{r_d},\,
\frac{D_H(z)}{r_d},\,
\frac{D_V(z)}{r_d}
\right\}_{\!\!{\rm th}},
\end{equation}
where, we fixed to the 
$\Lambda$CDM value $r_d = 147.09\,{\rm Mpc}$ from Planck \cite{Planck:2018vyg}.

Then, to perform the Bayesian analysis, we employ the Markov Chain Monte Carlo (MCMC) method to estimate the posterior distributions of the model parameters using the publicly available package \textbf{emcee} \cite{Foreman_Mackey_2013}. The convergence of the MCMC chains is analyzed with \textbf{GetDist} \cite{Lewis:2019xzd}. 

\subsection{Interaction function 
$f(t) = \beta \rho_m(t)$}
\begin{figure}[!h]
    \centering
        \includegraphics[scale=0.5]{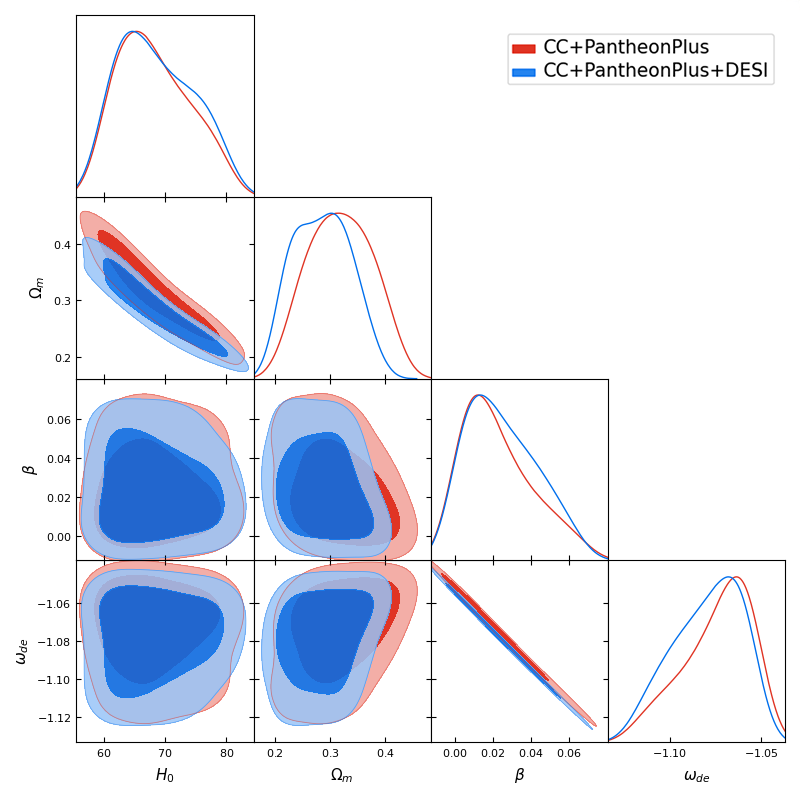}
        \caption{Constraints on parameters $\{H_0, \Omega_m, \beta, \omega_{de}\}$ for the first interaction $f\propto \rho_m$. }
    \label{FIG_contour1}
\end{figure}

In this section, we present the marginalized constraints on the cosmological parameters $\{H_0, \Omega_m, \beta, \omega_{de}\}$ at the 68\% and 95\% confidence levels for an interaction function proportional to the matter energy density, $f(t) = \beta \rho_m(t)$, as shown in Table \ref{mean_inter1} and Figure \ref{FIG_contour1}. It is also important to note that, from the critical points analysis in the previous section for this interaction function, we require $\alpha = -1$ in order to obtain a matter-dominated era, thereby constraining this parameter.

\begin{table}[h!]
\centering
\caption{Marginalized constraints on the cosmological parameters at 68\% and 95\% confidence levels for $f(t) = \beta \rho_m(t)$.}
\begin{tabular}{ccc}
\hline \hline
\textbf{Parameter} & \textbf{\ \ \ \ \ \ \ CC + PantheonPlus \ \ \ \ \ \ \ } & \textbf{CC + PantheonPlus  + DESI} \\
\hline
$H_0$ & $68.20^{+4.88+11.93}_{-7.66-10.40}$ & $68.54^{+5.43+12.10}_{-7.99-10.79}$ \\[4pt]
$\Omega_m$ & $0.319^{+0.059+0.096}_{-0.059-0.094}$ & $0.286^{+0.051+0.098}_{-0.056-0.096}$ \\[4pt]
$\beta$ & $0.0227^{+0.0126+0.0404}_{-0.0245-0.0319}$ & $0.0248^{+0.0152+0.0374}_{-0.0243-0.0329}$ \\[4pt]
$w_{de}$ & $-1.074^{+0.0243+0.0326}_{-0.0132-0.0399}$ & $-1.079^{+0.0244+0.0333}_{-0.0164-0.0378}$ \\[4pt]
\hline\hline
\end{tabular}
\label{mean_inter1}
\end{table}

\subsection{Interaction function 
$f(t) = \beta \rho_{de}(t)$}

\begin{figure}[!h]
    \centering
        \includegraphics[scale=0.5]{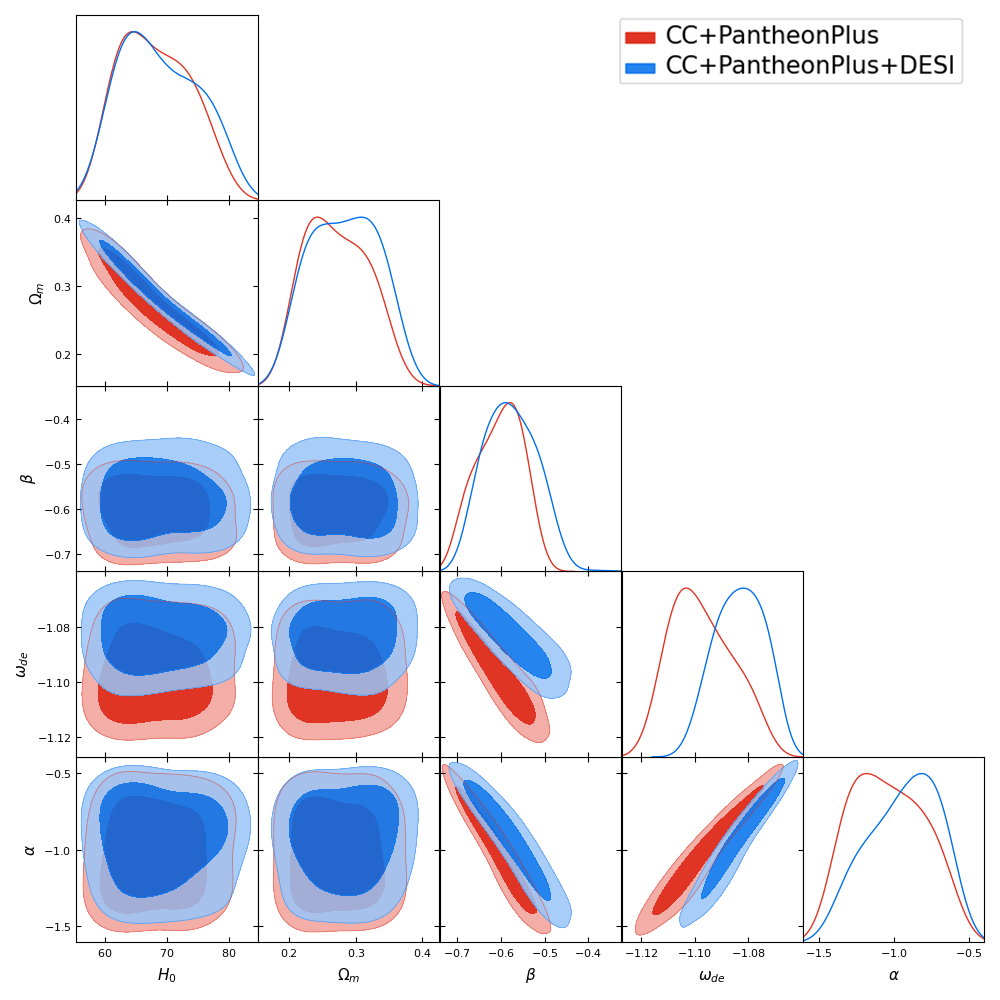}
        \caption{Constraints on parameters $\{H_0, \Omega_m, \beta, \omega_{de}, \alpha\}$. }
    \label{FIG_contour2}
\end{figure}

In this section, we present the marginalized constraints on the cosmological parameters $\{H_0, \Omega_m, \beta, \omega_{de}, \alpha\}$ at the 68\% and 95\% confidence levels for an interaction function proportional to the density of matter energy, $f(t) = \beta \rho_{de}(t)$, as shown in Table \ref{mean_inter2} and Figure~\ref{FIG_contour2}. 

\begin{table}[h!]
\centering
\caption{Marginalized constraints on the cosmological parameters at 68\% and 95\% confidence levels for $f(t) = \beta \rho_{de}(t)$.}
\begin{tabular}{ccc}
\hline \hline
\textbf{Parameter} & \textbf{\ \ \ \ \ \ \ CC + PantheonPlus \ \ \ \ \ \ \ } & \textbf{CC + PantheonPlus  + DESI} \\
\hline
$H_0$ & $68.17^{+5.41+11.10}_{-7.17-10.32}$ & $68.90^{+5.80+12.23}_{-8.07-11.07}$ \\[4pt]
$\Omega_m$ & $0.272^{+0.046+0.090}_{-0.058-0.085}$ & $0.283^{+0.053+0.093}_{-0.056-0.094}$ \\[4pt]
$\beta$ & $-0.603^{+0.065+0.095}_{-0.050-0.104}$ & $-0.577^{+0.060+0.114}_{-0.068-0.111}$ \\[4pt]
$w_{de}$ & $-1.097^{+0.010+0.024}_{-0.015-0.021}$ & $-1.083^{+0.011+0.018}_{-0.009-0.018}$ \\[4pt]
$\alpha$ & $-1.037^{+0.238+0.440}_{-0.304-0.440}$ & $-0.935^{+0.303+0.467}_{-0.205-0.467}$ \\[4pt]
\hline\hline
\end{tabular}
\label{mean_inter2}
\end{table}

\subsection{Interaction function $f(t) = \beta [\,\rho_m(t)+\rho_{de}(t)\,]$}

\begin{figure}[!h]
    \centering
        \includegraphics[scale=0.5]{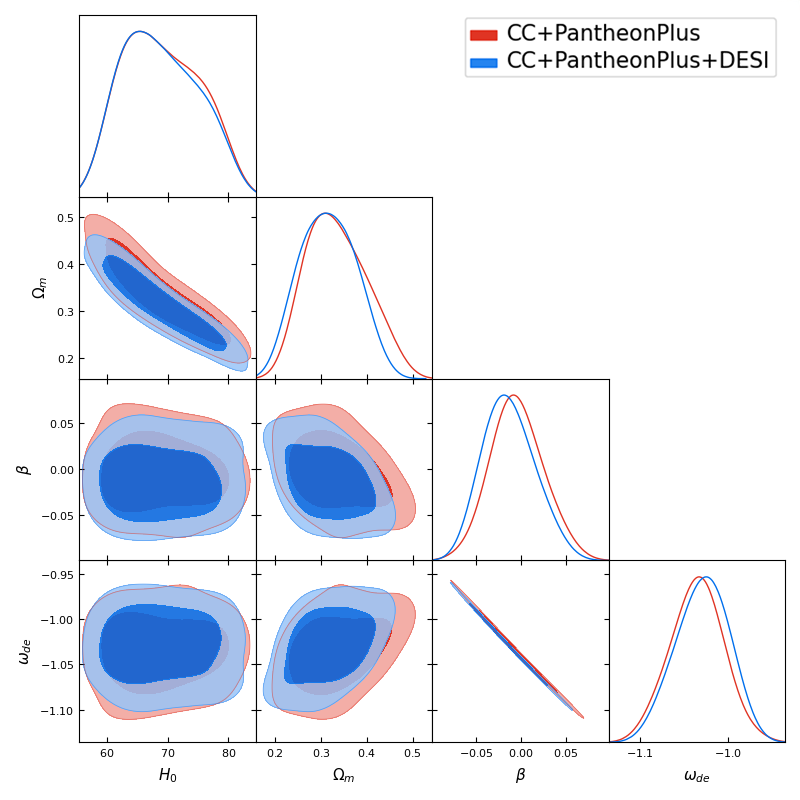}
        \caption{Constraints on parameters $\{H_0, \Omega_m, \beta, \omega_{de}\}$. }
    \label{FIG_contour3}
\end{figure}

In this section, we present the marginalized constraints on the cosmological parameters $\{H_0, \Omega_m, \beta, \omega_{de}\}$ at the 68\% and 95\% confidence levels for an interaction function proportional to the matter energy density, $f(t) = \beta [\,\rho_m(t)+\rho_{de}(t)\,]$, as shown in Table \ref{mean_inter3} and Figure~\ref{FIG_contour3}. It is also important to note that, from the critical points analysis in the previous section for this interaction function, we require $\alpha = -1$ in order to obtain a matter-dominated era, thereby constraining this parameter.

\begin{table}[h!]
\centering
\caption{Marginalized constraints on the cosmological parameters at 68\% and 95\% confidence levels for $f(t) = \beta [\,\rho_m(t)+\rho_{de}(t)\,]$.}
\begin{tabular}{ccc}
\hline \hline
\textbf{Parameter} & \textbf{\ \ \ \ \ \ \ CC + PantheonPlus \ \ \ \ \ \ \ } & \textbf{CC + PantheonPlus  + DESI} \\
\hline
$H_0$ & $68.98^{+5.90+11.90}_{-7.76-11.03}$ & $68.77^{+5.56+11.95}_{-7.82-10.88}$ \\[4pt]
$\Omega_m$ & $0.335^{+0.059+0.133}_{-0.081-0.120}$ & $0.315^{+0.060+0.116}_{-0.060-0.115}$ \\[4pt]
$\beta$ & $-0.0045^{+0.0281+0.0560}_{-0.0320-0.0511}$ & $-0.013^{+0.027+0.059}_{-0.033-0.055}$ \\[4pt]
$w_{de}$ & $-1.035^{+0.032+0.054}_{-0.030-0.056}$ & $-1.029^{+0.034+0.058}_{-0.029-0.060}$ \\[4pt]
\hline\hline
\end{tabular}
\label{mean_inter3}
\end{table}

\section{Conclusions}\label{conclusion}

In this article, we examine a late-time interacting model within the framework of generalized Rastall gravity. In this theoretical construct, the interaction emerges naturally from the non-conservation of the energy-momentum tensor. Within this context, we analyze the background evolution of an interacting dark sector as an autonomous system derived from the dynamical variables. Concerning the interaction terms present in the equation of motion related to dark matter and dark energy, we have defined these interactions as $Q_1=\alpha\dot{f}$ and $Q_2=-\dot{f}(1+\alpha)$, with $\alpha$ a constant parameter. In order to study different types of interactions,  we have analyzed  the three simplest functions $f(t)$, given by  $f\propto \rho_m$, $f\propto \rho_{de}$ and the sum of both densities $f\propto \rho_m+\rho_{de}$, assuming constant the EoS parameter related to the dark energy $w_{de}$. 

In relation to the different interactions, given by the three  functions $f(t)$, we have  obtained a closed set for dynamical variables $(\Omega_{ de},\Omega_m)$.  In this context, we have  identified the fixed points  and we have also determined their stability properties over the relevant parameter domain. From the phase–space,  we have obtained  an unstable radiation point  that guarantees exit from the early epoch (radiation), an unstable critical point related to the matter-dominated epoch, and a late–time dark–energy critical point  associated  to a stable attractor, see Figs.\ref{Fig1},\ref{Fig3} and \ref{Fig5}.

In addition, we have also constrained the free parameters of each  interaction-model, using  a joint likelihood analysis based on three independent cosmological probes;  Cosmic Chronometers, Type Ia Supernovae, and Baryon Acoustic Oscillations  from the DESI survey.

Concerning to the first interaction defined as $f=\beta\rho_m$, we have found that to obtain an unstable critical point $b_M$ associated to the matter-dominated era is necessary that the parameter $\alpha$ must the value $\alpha=-1$. Thus, we have obtained that this condition implies that the model shows an interaction only with dark matter, since the interaction source related  with dark energy is zero, i.e., $Q_2 = 0$, (see Eqs.(\ref{Q2})). In this sense, we have obtained that the interaction term is   present only in the matter equation of motion, and not in the dark energy component, and this mechanism   may be associated with particle production. In relation to the Rastall parameter $\lambda_\text{Ras}(z)$ and the coupling function $g(z)$, we have found that for the inferior marginalized values of the parameters, both functions $g(z)$  and the generalized Rastall parameter $\lambda_{\rm Ras}(z)$ remain close to unity over an extended redshift range and these functions approach the GR limit. The superior and mean marginalized values present a small difference in these function for redshift $z>10$. However, we have found that in the future ($z<0$) and considering superior and mean marginalized values,  these functions $g(z)$ and $\lambda_\text{Ras}(z)$ approach to the unity (GR limit), see Fig.\ref{Fig2}. In relation to the marginalized constraints on the parameters, we have found that comparing the CC+PantheonPlus+DESI and CC+PantheonPlus data, the mean  value  Hubble parameter $H_0$ presents a small increase when we include the DESI results. The same situation occurs for the parameter $\beta$, and moreover where the mean value found of $\beta$  is positive. For the other two parameters $\Omega_m$ and $w_{de}$, we have obtained the result that the inclusion of DESI data decreases the mean values of these parameters, see Table \ref{mean_inter1}.   
 
Regarding the second interaction function $f\propto\rho_{de}$, we have found that the matter-dominated epoch does not impose a restriction on the parameter $\alpha$. In this context, we have obtained that the stability analysis of the critical points associated with  the second interaction shows that the system approaches a late-time attractor (dark energy) as plotted in Fig.\ref{Fig3}, where the stable critical point corresponds to the point $c$. 
In relation to the functions $g(z)$ and $\lambda_\text{Ras}(z)$, as a function of the redshift, we find that these quantities remain nearly constant and close to unity for redshifts $z > 10$. Moreover, this behavior is essentially independent of the specific values of the model parameters. However, as shown in Fig.\ref{Fig4}, for redshifts $z < 9$, and in particular at the present epoch, we find that this interaction model departs significantly from the GR case when we consider the best-fit values of our parameters at 1-$\sigma$ from the CC + PantheonPlus + DESI combination.

In relation to the marginalized constraints on the parameters for our second interaction in which $f\propto\rho_{de}$, we have found that comparing the CC+PantheonPlus+DESI and CC+PantheonPlus data, the mean  value  Hubble parameter $H_0$ presents a small increase when we incorporate the DESI results. The same situation takes place for the parameter $\beta$ and moreover we have obtained that  mean value  is negative and around the value of $-0.6$. For the other two parameters $\Omega_m$ and $w_{de}$, we have determined  that the inclusion of  DESI data increases the mean values of these parameters. Also, as in the first interaction, we have observed that the mean value of the EoS parameter is associated with dark energy $w_{de}<-1$  using CC+PantheonPlus  and $-1<w_{de}$ by using DESI results. In relation to the parameter $\alpha$, we have determined  that the mean value obtained is $-1.037$ using CC+PantheonPlus and $-0.935$ by adding DESI data (see Table \ref{mean_inter2}).

 Completing the study  we have considered the interaction term $f(t)=\beta\,[\rho_m(t)+\rho_{de}(t)]$. For this interaction, we have determined  that to obtain an unstable critical point $j_M$ related to the matter-dominated era  is necessary that the parameter $\alpha$ must the value  $\alpha=-1$, as in the case of the first interaction model studied. 
 In relation to the parameter $g$, as a function of the redshift, we have found   that this parameter  remains nearly constant for large values of the redshift ($z>10^4$).
 We have also obtained that the function $g$ does not approach the GR value (where $g = 1$). In addition,  from the lower panel of Fig.\ref{Fig6}, we have    noted that the Rastall parameter $\lambda_\text{Ras}$ remains approximately constant for redshift values $z>30$.
 Also, from this panel we have observed that the Rastall parameter $\lambda_\text{Ras}(z)$ does not approach to GR limit for values of the  redshift $z\sim 0$. In this context, we have found that  for the interaction $f\propto \rho_m+\rho_{de}$, the generalized Rastall gravity presents a slight deviation in relation to GR theory using the values of the parameters at 1-$\sigma$, from Table \ref{mean_inter3}.
 In relation to  the marginalized constraints on the parameters described by the Table \ref{mean_inter3}, we have found  that comparing the CC+PantheonPlus+DESI and CC+PantheonPlus data, the mean  value  Hubble parameter $H_0$ presents a small decrease when we include DESI data. In relation to the interaction parameter $\beta$  we have  found that the  mean value  is negative, as in the first interaction function. For the other two parameters $\Omega_m$ and $w_{de}$, we find that the inclusion of the DESI data decreases the mean values of the density parameter $\Omega_m$ and increases the value of EoS parameter $w_{de}$. As in the first interaction model, we have  observed  that the EoS parameter associated with the dark energy $w_{de}<-1$, presents a phantom behavior.

Finally, we highlight that the interacting model developed within the framework of generalized Rastall gravity opens several promising avenues for future research. The present analysis naturally suggests extending the study to alternative interaction schemes defined through the function $f(t)$, as well as exploring the consequences of these interactions for the growth of cosmic structures. These lines of research offer great potential to deepen our understanding of non-conservative gravitational dynamics within the framework of generalized Rastall gravity, and we intend to address them in future work.

\begin{acknowledgments}
M.G.-E. acknowledges the financial support of FONDECYT de Postdoctorado, No. 3230801. C.R.-B. acknowledges the financial support of Grant No. PE501082885-2023-PROCIENCIA. 

\end{acknowledgments}

\bibliography{bio}




\end{document}